\documentclass[twocolumn,tightenlines,prd,floatfix,nofootinbib]{revtex4}

 \usepackage[dvips,final]{graphicx}
  \usepackage{amssymb}
   \usepackage{amsmath}
    \usepackage{amsfonts}
     \usepackage{epsfig}
      \usepackage{bm}
\usepackage{slashed}

\newcommand{\GeV}{\textrm{GeV}}

\newcommand{\ie}{{i.e.}}
\newcommand{\eg}{{e.g.}}

\def\acal{\relax\ifmmode{\cal A}\else{${\cal A}${ }}\fi}

\begin{document}

\title{QCD rescattering mechanism for diffractive deep inelastic scattering}

\author{Roman Pasechnik}
\email{roman.pasechnik@fysast.uu.se}
\author{Rikard Enberg}
\email{rikard.enberg@physics.uu.se}
\author{Gunnar Ingelman}
\email{gunnar.ingelman@physics.uu.se}

\affiliation{Department of Physics and Astronomy, Uppsala
University, Box 516, SE-751 20 Uppsala, Sweden}

\begin{abstract}
We present a QCD-based model where 
rescattering between final state partons in deep inelastic
scattering leads to events with large rapidity gaps and a leading
proton. In the framework of this model the amplitude for multiple
gluon exchanges is calculated in the eikonal approximation to all
orders in perturbation theory. Both large and small invariant mass
$M_X$ limits are considered. The model successfully describes the 
precise HERA data on the diffractive deep inelastic cross section 
in the whole available kinematical range and gives new insight into 
the density of gluons at very small momentum fractions in the proton.
\end{abstract}


\maketitle

\section{Introduction}

Hadronic processes with a hard scale involved constitute an 
indispensable tool for probing the QCD dynamics of
quarks and gluons, and through the QCD factorization theorems \cite{QCDfac} that separate 
physics at small and large distances, one may also study the dynamics of soft processes 
with small momentum transfers. Hard quark and gluon interactions at small distances are
thus not affected by soft interactions and are described in perturbative QCD. 
The most problematic part of the process
is soft interactions at large distances, where nonperturbative QCD comes
into the game and manifests itself as
the confinement of quarks and gluons in hadrons and the related
hadronization process giving the observable hadronic final states in
high energy collisions.

Diffractive processes are sensitive
to the details of nonperturbative QCD dynamics and provide a way 
to probe the soft and semihard
regimes directly. Diffractive events are characterized by a leading ``target'' particle,
carrying most of the beam momentum, and a well separated produced hadronic
system. The ``gap'' in between is connected to the soft part of the
event and therefore to nonperturbative effects at a long
space-time scale. Diffractive deep inelastic
scattering (DDIS) offers a particularly good opportunity to explore the
interplay between hard and soft physics due to the precise
data from the electron--proton collider HERA
\cite{Chekanov:2008fh,ddisexp}.

DDIS in lepton--proton collisions
involves hard scattering events where, in spite of the large momentum transfer 
$Q^2$ from the electron, the proton emerges essentially unscathed with 
small transverse momentum, keeping almost all of its original 
longitudinal beam momentum (for reviews on DDIS, see e.g. 
Refs.~\cite{Hebecker99,WM99,Ingelman:2005ku}). 
The leading proton is well separated in momentum space, or rapidity $y=1/2 \, \ln(E+p_z)/(E-p_z)$, 
from the central hadronic system 
produced from the exchanged virtual photon's interaction with the proton. Thus, 
this new class of events is characterized by a {\em large rapidity gap} 
(LRG) void of final state particles.

Rapidity gaps in deep inelastic scattering (DIS) were discovered by 
the ZEUS and H1 experiments at HERA \cite{ddisexp}, but the first discovery 
of hard diffraction was in $p\bar{p}$ collisions by the 
UA8 experiment \cite{UA8}. These processes had actually been predicted 
\cite{IS85} by combining Regge phenomenology for soft 
processes in strong interactions via pomeron exchange, with hard processes 
based on perturbative QCD. By parametrizing the 
parton content of an exchanged pomeron (or alternatively diffractive 
parton density functions) it is possible to describe the HERA data. 
However, the extracted parton densities are not universal, since when 
used to calculate diffractive hard scattering processes in $p\bar p$ 
collisions at the Tevatron one obtains cross sections an order of 
magnitude larger than observed.
\begin{figure}[b]    
 \centerline{\includegraphics[width=0.45\textwidth]{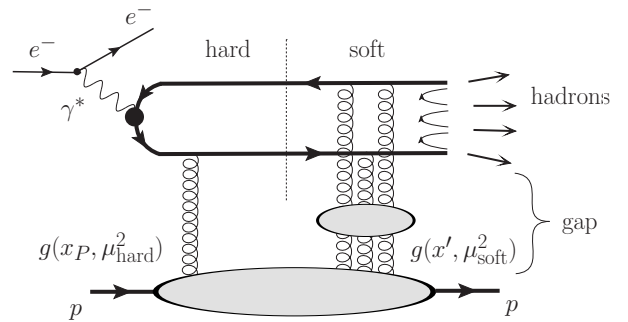}}
   \caption{\label{fig:DDIS-cartoon}
   Schematic illustration of the mechanism for diffractive deep inelastic scattering considered in this paper, with 
   soft gluon exchanges in the final state.}
\end{figure}

An alternative dynamical interpretation of hard diffraction was proposed in Refs.~\cite{Edin:1995gi}. 
This Soft Color Interaction
(SCI) model is based on the simple assumption of soft gluon exchanges leading to color rearrangements
between the final state partons. Variations in the topology of the
confining color fields lead to different hadronic final states. 

The SCI model is implemented in Monte Carlo event generators, 
\eg\ {\sc Lepto} for DIS \cite{Ingelman:1996mq}.
The hard part of the process shown in Fig.~\ref{fig:DDIS-cartoon} 
is then calculated in the framework of perturbative QCD with DGLAP
evolution of the parton showers in the same way as in inclusive DIS. 
The large momentum transfer means that the hard subprocess occurs on a 
spacetime scale much smaller than the bound state proton and is thus 
``embedded'' in the proton. The emerging hard-scattered partons 
propagate through the proton's color field and may interact 
with it. Soft exchanges will dominate due to the large coupling and 
the lack of suppression from hard gluon propagators. Therefore, 
the momenta of the hard partons are essentially undisturbed --- 
the soft, long distance interactions do not affect the hard, short distance process, and the 
momentum transfer of the soft exchanges can be neglected. 
However, the exchange of color changes the color charges of the emerging partons 
such that the confining string-like field between them will have a different topology, 
resulting in a different distribution of final state hadrons produced from the 
string hadronization~\cite{Andersson:1983ia}. In particular, a region in rapidity 
without a string will result in an absence of hadrons there, i.e.\ a rapidity gap.

The only parameter of this model is the probability for a soft exchange,
accounting for the unknown nonperturbative dynamics. Remarkably, the SCI model is
phenomenologically very successful in describing many different processes, both diffractive and nondiffractive \cite{sciresult}, with only a single parameter $P\simeq 0.5$ for this probability. 
Thus, the SCI model captures the essential dynamics of diffraction,
but lacks a solid theoretical basis. 

To understand better what we can learn from the phenomenology of the SCI model, we
present in this paper a detailed QCD-based mechanism for soft gluon rescattering of 
final state partons, as 
illustrated in Fig.~\ref{fig:DDIS-cartoon}. This mechanism 
leads to effective color singlet exchange and thereby to diffractive scattering. 
Inspired by the SCI model, the model presented here may be seen as an explicit realization 
of the earlier attempt \cite{BEHI05} to understand soft gluon exchange 
in terms of QCD rescattering. Our model was initially introduced
in a recent letter \cite{our-hep}, and is here presented in detail.

The paper is organized as follows. In Section II we briefly discuss the 
framework of the dipole approach and motivate our study. 
In Section III we consider the kinematics of diffractive DIS. Section IV treats the formalism 
for generalized unintegrated gluon distribution functions in the diffractive 
limit. The explicit calculation of the $q{\bar q}$ dipole contribution to 
the diffractive cross section and analytic approximations 
used are presented in Section V. In Section VI we study the contribution 
of the $q{\bar q}g$ final state. Numerical results and comparisons with HERA data 
on the diffractive cross section are given in Section VII. 
Finally, in Section VIII we present some concluding remarks and an outlook.

\section{Dipole approach}

\begin{figure}[tb]    
 \centerline{\includegraphics[width=0.45\textwidth]{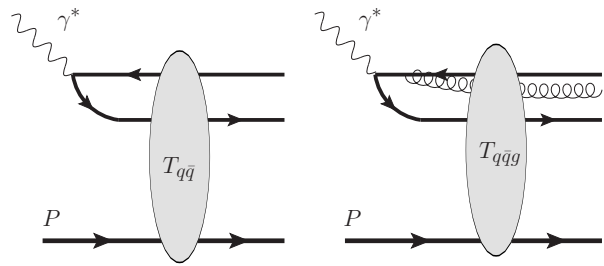}}
   \caption{\label{fig:DDISa}
   Typical diagrams contributing to diffractive DIS with the leading order $q{\bar q}$ dipole scattering (left) and the gluonic $q{\bar q}g$ contribution (right) contributing significantly at $\beta\to0$ (large $M_X$).}
\end{figure}

Typical contributions to the diffractive DIS process are represented
by the diagrams in Fig.~\ref{fig:DDISa}. In terms of the four-momenta 
$q$ of the photon, and $P$ and $P'$ of the initial and final proton, 
the kinematics of the $\gamma^*P\to X P'$ process is defined by the variables
 \begin{equation}\label{vars}
 x_B=\frac{Q^2}{Q^2+W^2}\,,\quad \beta=\frac{Q^2}{Q^2+M_X^2}\,,\quad
 x_P=\frac{x_B}{\beta}\,,
 \end{equation}
where $Q^2=-q^2$. The invariant mass of the produced system $M_X$, and the total
energy in the $\gamma^*P$ center-of-mass system $W$ are given by
 \begin{equation}\label{vars1}
  M_X^2=\frac{1-\beta}{\beta}Q^2\,,\quad
 W^2\equiv (P+q)^2 = \frac{Q^2}{x_B}(1-x_B)\,.
 \end{equation}
The DDIS cross section in general
is represented as a function of $\beta,\,x_P,\,Q^2$ and the momentum
transfer along the proton line $t=(P'-P)^2$. Note, that we are
working in the forward limit of small $|t|\ll Q^2,\,M_X^2$.

\begin{figure}[b]
 \centerline{\includegraphics[width=0.4\textwidth]{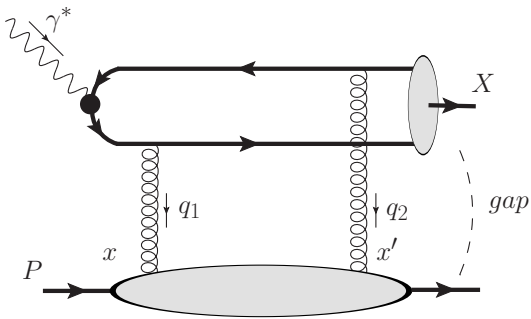}}
   \caption{\label{fig:ampLO}
   Amplitude of the process $\gamma^*p\to Xp$
   at the leading order of perturbation theory.}
\end{figure}

Let us consider first the simplest case of the $q{\bar q}$ contribution, 
which is the leading one for small $M_X$ (or, equivalently, $\beta\to 1$). 
To compute the
diffractive DIS amplitude, it is convenient to consider the process
in the dipole frame \cite{dipole}, where the deeply virtual photon
with large virtuality $Q^2$ and polarization $\lambda$ first
splits into a quark $q$ and an antiquark ${\bar q}$ with mass $m_q$,
spins $\alpha$ and $\beta$, and flavor $f$, and then the $q{\bar q}$
dipole with transverse size $r$ interacts with the target proton
at impact parameter $b$ and dissociates into a final state $X$ of
invariant mass $M_X$ as shown in Fig.~\ref{fig:ampLO}. The photon
splitting into the dipole is a QED process and is described by the 
wavefunctions $\psi^{f,\alpha\beta}_{\lambda}(z,{\bf r};Q^2)$ in the impact
parameter space \cite{dipole,ddiswf}, where $z$ is the fraction of the
longitudinal momentum carried by the quark. The amplitude of the
dipole--nucleon scattering (denoted as $T$ in Fig.~\ref{fig:DDISa})
is the only unknown non-perturbative object (for a review, see
Ref.~\cite{WM99}). The dipole picture naturally 
incorporates the description of both inclusive and diffractive events 
into a common theoretical framework~\cite{nikzak,ddiswf}, as the same 
dipole scattering amplitudes enter into inclusive and diffractive 
cross sections.

Final states with gluons are suppressed by powers of $\alpha_s$.
However, if $\beta$ becomes small or $Q^2$ large, the $q{\bar q}$
dipole emits soft or collinear gluons accompanied by large
logarithms $\ln(1/\beta)$ or $\ln Q^2$ which compensate the
suppression in $\alpha_s$ \cite{gwus}. The $q\bar{q}$ pair can emit
soft gluons, leading to the dressing up of the quarks, which is
parametrized by a scale-dependent constituent quark mass
$m^{\mathrm{eff}}_q(\mu^2)$. In general, the more gluons in the
final state, the larger the invariant mass produced. The dominant gluon
emission from quarks is described by DGLAP evolution \cite{DGLAP} and 
is mostly collinear to the radiating quark, so it cannot build up a large 
$M_X$. The small $\beta\to0$ and large $Q^2\to \infty$ limits can be driven,
therefore, only by a semihard gluon radiation from the active gluon
(carrying $x_P$) giving rise to a gluonic dipole $q\bar qg$ contribution.
These aspects will be discussed in detail below.

Diffractive DIS at the leading order in $\alpha_s$ is described by
the two gluon exchange diagram shown in Fig.~\ref{fig:ampLO}. Let
us first discuss how the longitudinal momentum transfer between the
$q\bar{q}$ dipole and the proton can be shared between the gluons.
The gluon momenta can be Sudakov decomposed as
 \begin{equation} \nonumber
 q_1=-xP+\Delta_{\perp}\,,\quad q_2=-x' P+\Delta'_{\perp}\,,\quad
 x+x'=x_P\,.
 \end{equation}
Cutting the diagram after the first gluon exchange (picking up the
leading poles only), in the high-energy limit $x_B\to 0$, we have
 \begin{equation} \nonumber
 (q+xP-\Delta_{\perp})^2=M_\text{int}^2\;\to\;-2Pq(x_B-x)=M_{\text{int},\perp}^2
 \end{equation}
In the deep inelastic limit $Q^2\to\infty$, for fixed invariant mass
of the intermediate system $M_\text{int}\sim M_X$ and Bjorken variable
$x_B$, we see that $M_\text{int}^2\ll 2\,Pq$ and $x\simeq x_B$, thus
$x'=x_P-x_B$. On the other hand, when $Q^2\to\infty$ and $M_\text{int}^2$
fixed, we see that $\beta=x_B/x_P\to1$. So $x\to x_P$ and $x'\to
0$, and the first gluon takes all the longitudinal momentum exchange
neutralizing the virtuality of the $q\bar{q}$ system.
The latter kinematical
configuration gives the leading contribution to the cross section,
whereas the other configurations with equal momentum sharing between
the gluons $x\sim x'\sim x_P/2$ are suppressed by extra propagators.

\begin{figure}[tb]
 \centerline{\includegraphics[width=0.4\textwidth]{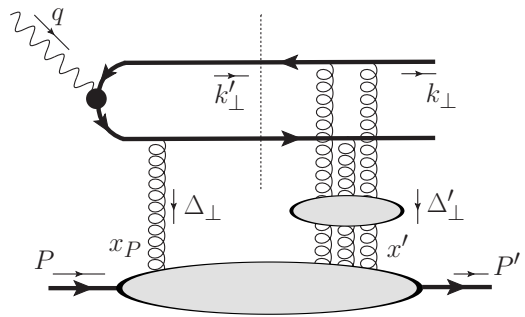}}
   \caption{\label{fig:amp}
   Amplitude of the process $\gamma^*p\to Xp$
   with all-order
   resummed soft gluon exchange.}
\end{figure}

Thus, we consider the asymmetric case with one hard
(perturbative) gluon carrying most of the longitudinal momentum
transfer $x_P$, and a number of multiple soft screening interactions
with total $x'\ll x_P$ in a color octet state (in the large $N_c$
limit) effectively described by the resummed multigluon exchange
amplitude as schematically sketched in Fig.~\ref{fig:amp}. The
first ``hard'' gluon turns the proton into a color octet proton remnant,
which then interacts with a system of soft screening gluons in an
octet state and finally recombines into a color singlet
corresponding to the leading proton (or system with invariant mass
close to the proton mass). These soft gluons cannot dynamically resolve a
$q\bar q$ dipole of small size $r\sim 1/k_\perp$, However, we assume that 
the gluons interact with the quark current and not with the dipole as a whole. 
This will be further discussed below in Sec.~\ref{interpretation}.

\section{Kinematics of diffractive DIS}

Let us first define the kinematics of the process 
$\gamma^*p\to X p$. 
Our primary interest is processes with small momentum
transfer $t\ll Q^2$. It is convenient to fix the
frame of reference in the center-of-mass system of the final states, i.e.\ 
the outgoing proton with momentum $P'$, and the diffractive system
$X$ with momentum $q'=k_1+k_2$, ${\bf k}_{1,\perp}=-{\bf
k}_{2,\perp}\equiv {\bf k}_{\perp}$.

The total $\gamma^*p$ c.m.s.\ energy squared is
$W^2=(P'+q')^2=2P'q'+M_X^2$ (the proton mass $m_p$ is neglected). In
terms of the cross section variables defined in Eqs.~(\ref{vars}) and (\ref{vars1}) 
we write
\begin{eqnarray} \nonumber
2P'q'=W^2-M_X^2=Q^2\frac{1-x_P}{x_B},\quad
2Pq=\frac{Q^2}{x_B}\simeq 2P'q'.
\end{eqnarray}
Since $x_P\ll 1$, we have $W^2\gg M_X^2$ for any $\beta$ and $Q^2$, so we will first keep
$M_X^2$ and then drop it in comparison with $W^2$ whenever appropriate.

The general Sudakov decompositions of the final quark/antiquark
momenta $k_{1,2}$ are
\begin{eqnarray} \nonumber
k_1=(1-z)q'+n_1P'+k_{1,\perp},\quad k_2=zq'+n_2P'+k_{2,\perp}\,.
\end{eqnarray}
where $n_1=-n_2$. The on-shell conditions for the quark and antiquark in the
final state
\begin{eqnarray*}
&&(1-z)^2M_X^2+n_1(1-z)(W^2-M_X^2)-k_{\perp}^2=m_q^2,\\
&&\quad z^2M_X^2+n_2z(W^2-M_X^2)-k_{\perp}^2=m_q^2,
\end{eqnarray*}
give
\begin{eqnarray} \nonumber
n_1=\frac{m_{q,\perp}^2-(1-z)^2M_X^2}{(1-z)(W^2-M_X^2)},\quad
n_2=\frac{m_{q,\perp}^2-z^2M_X^2}{z(W^2-M_X^2)}.
\end{eqnarray}
Finally, $n_2=-n_1\equiv n$ leads to
\begin{eqnarray} \label{mn}
M_X^2=\frac{m_{q,\perp}^2}{z(1-z)},\quad
n=(1-2z)\frac{M_X^2}{(W^2-M_X^2)}
\end{eqnarray}
where $m_{q,\perp}^2=m_q^2+k_{\perp}^2$ is the transverse quark mass
squared. Applying Eq.~(\ref{vars1}), we get the standard relation
for the quark transverse momentum
\begin{eqnarray} \label{kt}
k_{\perp}^2=z(1-z)M_X^2-m_q^2\,,
\end{eqnarray}
and we consider the light quark mass limit $m_q\ll M_X^2$. 
The leading contribution to diffractive DIS at HERA
comes from light quarks, and from now on we do
not distinguish between their masses and account for them by one
single mass parameter $m_q^{\text{eff}}$.

In the diffractive limit $Q^2,\,M_X^2\ll W^2$, 
when $z$ and $1-z$ are not very asymmetric, one has with good accuracy:
\begin{eqnarray*}
k_1^0\simeq k_1^z \simeq (1-z)\frac{W}{2},\quad k_2^0\simeq k_2^z
\simeq z\frac{W}{2}\,,
\end{eqnarray*}

Let us now define the quark propagators in the photon
splitting wave function. Due to the condition $x_P\gg x_{2,3,\, \dots}$ the
soft screening gluons cannot change the longitudinal momenta
significantly, but only the transverse momenta. Thus, to calculate the hard
part of the amplitude let us first neglect these extra
screening gluons. We will show below that adding the extra soft
gluons leads only to phase shifts (and their derivatives) in the 
transverse coordinates, which are going be resummed to all orders 
in $\alpha_s$.

In the chosen frame, the momenta of the
exchanged hard and the sum of the soft gluons are
\begin{eqnarray} \nonumber
q_1\simeq-x_P P'+\Delta_{\perp},\;\; q_2\simeq \Delta'_{\perp},\;\;
\delta=|{\bm \Delta}'_{\perp}+{\bm
\Delta}_{\perp}|\simeq\sqrt{-t}\,.
\end{eqnarray}
Let us first attach the hard gluon to the lower quark line $k_2$. Then
the denominator of the quark propagator between the photon and gluon vertices is
\begin{align}
(k_2+q_1)^2&-m_q^2 = \nonumber \\ 
&=-z^2M_X^2-zQ^2\frac{\beta-x_P}{\beta}-(k_{\perp}+\Delta_{\perp})^{2}-m_q^2\nonumber \\
&\simeq -\frac{\varepsilon^2+(1-z)(k_{\perp}+\Delta_{\perp})^{2}+zk_{\perp}^2}{1-z}\nonumber \\
&\simeq -\frac{\varepsilon^2+(k_{\perp}+\Delta_{\perp})^{2}}{1-z}
\label{pr1fin}
\end{align}
where $\varepsilon^2=z(1-z)Q^2+m_q^2$. The first approximation is 
obtained by substituting $M_X^2$ from Eq.~(\ref{mn}) in the limit 
$x_P\ll \beta$ and the second by using the limit $\Delta_{\perp}\ll k_{\perp}$, 
realized when $M_X\gg m_q,\Delta_{\perp}$, giving a result valid at $z\ll 1$.

When the gluon is attached to the upper gluon line, using momentum
conservation $q=q'+q_1+q_2$, we get analogously
\begin{eqnarray} \label{pr2fin}
(q-k'_2)^2-m_q^2\simeq-\frac{\varepsilon^2+(k_{\perp}-\Delta_{\perp})^2}{z},
\end{eqnarray}
which is strictly valid at $1-z\ll 1$. It is equal to
Eq.~(\ref{pr1fin}) with the exchanges $z\leftrightarrow (1-z)$ and
$k_{\perp}\leftrightarrow -k_{\perp}$.

The expressions (\ref{pr1fin}) and (\ref{pr2fin}) will be used for all values of $z$, 
as is common practice~\cite{WM99}. This is justified in our 
asymmetric case $x_P\gg x'$ because the dominating contribution to the amplitude 
comes from the configuration that either the quark or the antiquark from 
the photon is essentially on-shell, and the other carries the negative virtuality 
of the photon and then absorbs the hard gluon with momentum fraction $x_P$ to become essentially on-shell.

In the limit considered, $\Delta_{\perp}\ll k_{\perp}$, the quark
virtuality $k^2$ is conventionally utilized as the factorization scale $\mu_F^2$ 
of the process, and is expressed in terms of the energy $\varepsilon$ and the transverse
momentum $k_{\perp}$ as
\begin{eqnarray} \label{muF}
 \mu_F^2\equiv\varepsilon^2+k_{\perp}^2=z(1-z)(M_X^2+Q^2)\,.
\end{eqnarray}
Thus, the hard scale depends on both $Q^2$ from the space-like photon and $M_X^2$ from the 
time-like final state $X$. Since these can have any values, the QCD factorization is  
complicated and the physics may be different in the three cases $M_X\ll Q$, $M_X\sim Q$ and $M_X\gg Q$. 
The quark propagator (with the hard scale $\mu_F^2$ in Eq.~(\ref{muF})) is
antisymmetric, $k^2\leftrightarrow -k^2$, with respect to reflection
between the space-like and time-like regimes, i.e., with respect to the
interchange $M_X^2\leftrightarrow -Q^2$ (or
$\varepsilon^2\leftrightarrow -k_{\perp}^2$) as it should be.

The next step is to compute the bilinear spinor combinations $\bar
u(k_2,\lambda_q)\slashed{\epsilon}(\lambda_{\gamma})v(k_1,\lambda_{\bar
q})$ in the photon splitting $\gamma^*\to q{\bar q}$. 
The photon polarization vectors in the $XP'$ c.m.s.\ frame
have the following general form
\begin{eqnarray*}
&&\varepsilon^T_{\mu}(\lambda_{\gamma}=\pm1)=\frac{1}{\sqrt{2}}(0,\,1,\,\pm
i,\,0),\\
&&\varepsilon^L_{\mu}(\lambda_{\gamma}=0)=\frac{i}{2WQ}(W^2+Q^2,\,0,\,0,\,W^2-Q^2)\,.
\end{eqnarray*}
In particular, by straightforward calculation for the longitudinally
(L) polarized photon we simply get the following expression
\begin{eqnarray}
\bar
u_{\pm}(k_2)\slashed{\epsilon}(\lambda_{\gamma}=0)v_{\mp}(k_1)\simeq
i\sqrt{z(1-z)}Q\,. \label{Lexp}
\end{eqnarray}
which is not dependent on the transverse momenta of the initial quark
and antiquark.

The transversely polarized case requires a separate discussion. Within the dipole picture 
the diffractive DIS process can be basically decomposed into time-ordered
stages. First, the space-like photon with $q^2=-Q^2$ fluctuates into a
$q{\bar q}$ pair, which is then scattered off the target through 
hard gluon exchange, making the $q{\bar q}$-system time-like, and finally the on-shell quark and
antiquark scatter softly off the color background field in the
proton resulting in a color singlet $X$-system with invariant mass
$M_X$. Initially, at the moment of the photon fluctuation, the only
hard scale is $Q^2$, and the transverse momentum of a produced quark
is expressed through this scale as $k'_{\perp}\simeq
\sqrt{z(1-z)}Q$, which is different from the transverse momentum
$k_{\perp}$ of a quark in the final state defined in Eq.~(\ref{kt}).
In particular, the difference between $k_{\perp}$ and $k'_{\perp}$
can depend on the actual momentum transfer $\Delta_{\perp}$ for the
hard gluon and on the sum of the screening gluons, because according to
Eq.~(\ref{kt}) the small variation in $k_{\perp}$ due to the attached
$\Delta_{\perp}$ brings a significant change in $M_X$ for small
$z$. The relative coefficient between $k_{\perp}$ and $k'_{\perp}$ is
$Q/M_X$, and has to be taken into account when expressing the
transversely polarized photon splitting wave function through the
final state transverse momenta. This physical argument agrees well
with the kinematics corresponding to the diagram shown in
Fig.~\ref{fig:amp} for $Q\sim M_X$. In the opposite limits $Q\gg M_X$ or
$Q\ll M_X$ the emission of extra gluons
significantly complicates the kinematics, and this will be considered in detail below.

As a final result, for the transversely (T) polarized photon, we
get in the chiral limit
\begin{eqnarray}
\bar u_{\pm}(k_2)\slashed{\epsilon}(\lambda_{\gamma}=\pm1)v_{\mp}(k_1)=
\frac{Q}{M_X}\sqrt{\frac{2z}{1-z}}\,(k_1^x\pm i k_1^y) \,.
\label{Texp}
\end{eqnarray}
The spinor signs $\pm$ here stand for quark/antiquark chiralities,
which coincide with the helicity for a quark and have the opposite sign of the
helicity for an antiquark. We do not take into account the ``$++$'' and
``$--$'' components since they are small in the Bjorken
limit and for relatively light quark and antiquark.

\section{Generalized unintegrated gluon distributions}\label{sec-GPD}

\begin{figure}[bt]
 \centerline{\includegraphics[width=0.45\textwidth]{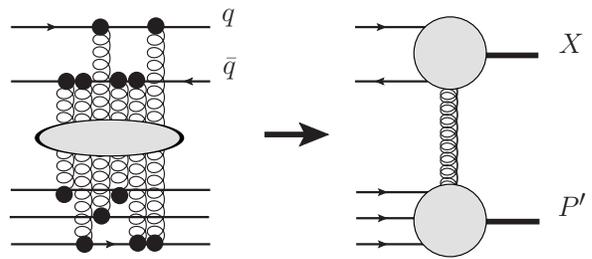}}
   \caption{\label{fig:smallxPDF}
   The sum of the multiple soft gluon exchanges can in an effective description be seen as a single gluon exchange between the $q{\bar q}$ dipole and the color octet proton remnant, resulting in the color singlet $X$-system and a proton in the final state.}
\end{figure}

Before calculating the amplitude for the hard and soft gluon exchanges, 
we note that the exchanged gluons all originate from the proton color field and should therefore 
be treated through a common description of a general gluon density. The first, hard 
gluon carries the dynamics through the longitudinal momentum $x_P P$, whereas the 
soft rescattering gluons carry small momenta $x_i'\ll x_P$ but may transfer a 
color octet charge that screens the color of the first gluon resulting in an overall 
color singlet exchange. In this sense, the sum of all soft exchanges acts 
as a single effective gluon exchange between the $q{\bar q}$ dipole and the proton remnant (see Fig.~\ref{fig:smallxPDF}).

As an appropriate description of the density of hard and soft effective gluons, 
we use the framework of generalized off-diagonal unintegrated gluon distribution 
functions (UGDF), which naturally appear in the $k_{\perp}$-factorization 
approach \cite{KT}. Within this framework 
the coupling to a quark is replaced by an off-diagonal UGDF 
${\cal F}_g^{\text{off}}(x_P,x',\Delta_{\perp}^{2},{\Delta'_{\perp}}^{2},\mu_F^2)$, 
absorbing a factor $C_F\alpha_s(\mu^2_{\text{soft}})/\pi$ and by convention also a gluon 
propagator $\Delta_\perp^{-2}$ in order to keep the UGDF regular as $\Delta_\perp^2\to 0$. 
The absorbed coupling $\alpha_s(\mu^2_{\text{soft}})$ corresponds to
the coupling of a screening gluon with virtuality
$\mu_{\text{soft}}^2\sim\Delta_{\perp}^2$ to a quark in the proton,
whereas the coupling of the hard gluon to the $q{\bar q}$ dipole is purely perturbative and occurs at the
hard scale $\mu_F$.

Generalized parton distributions (GPD) are not very constrained 
by data. We use a prescription for the generalized UGDF~\cite{Pasechnik:2007hm}, which
works well in the description of CDF data on central exclusive
charmonium production~\cite{Pasechnik:2009qc}. This prescription is motivated by positivity
constraints for the collinear GPDs \cite{Pire:1998nw} and can be considered 
as a saturation of the Cauchy--Schwarz inequality for the density matrix \cite{Artru:2008cp}. 
It incorporates the dependence on the longitudinal momentum fraction $x'$ 
and the transverse momenta of the soft gluons in an explicitly symmetric way,
 \begin{equation}
 {\cal F}^{\text{off}}_g\simeq
 \sqrt{ {\cal F}_g(x_P,\Delta_{\perp}^2,\mu_F^2)
 {\cal F}_g(x',{\Delta_{\perp}'}^2,\mu_{\text{soft}}^2)}\,, \label{sqrt}
 \end{equation}
with $\Delta_{\perp}\sim \Delta'_{\perp}$. Here ${\cal F}_g$ is the normal diagonal UGDF, 
which depends on the gluon virtuality $\Delta_{\perp}^2$ and reduces to the well-known collinear gluon PDF $g(x,\mu^2)$ when integrated over this virtuality.

This model (\ref{sqrt}) is a factorization of the
generalized UGDF into a hard part depending on a hard scale
$\mu_F$ and on $x_P$, thus describing the hard gluon coupling to the proton,
and a soft part defined at some soft scale $\mu_{\text{soft}}$ and
small $x'\ll x_P$ corresponding to a number of soft gluon
couplings. As we will see below, together with the factorization in
transverse momentum space, the model (\ref{sqrt}) provides a QCD
factorization of the diffractive amplitude in momentum
space.

The UGDF ${\cal F}_g(x,\Delta_{\perp}^2,\mu^2)$ introduced above depends on the
gluon virtuality, and this dependence is not theoretically
well-known for small virtualities. The UGDF is here modeled using the collinear gluon density 
$x_Pg(x_P,\mu_F^2)$, fixed at the QCD factorization
scale $\mu_F$, together with a simple Gaussian Ansatz for the dependence on the 
gluon virtuality $\Delta_{\perp}^2$ as
 \begin{align}
 \sqrt{x_P}{\cal F}^{\text{off}}_g &\simeq \sqrt{x_Pg(x_P,\mu_F^2)
 \, x'g(x',\mu_{\text{soft}}^2)}\,
 f_{G}(\Delta_{\perp}^2),\nonumber\\
 &f_{G}(\Delta_{\perp}^{2})={1}/({2\pi\rho_0^2})\,\exp\left({-{\Delta_{\perp}^2}/{2\rho_0^2}}\right),
 \label{ugdf}
 \end{align}
where the Gaussian width $\rho_0$ is the soft hadronic scale. As
demonstrated below, this scale corresponds to the transverse
proton size $r_p\sim 1/\rho_0$. The Gaussian smearing is then interpreted as the result of many soft interactions in the bound state proton. The factor $\sqrt{x_P}$ in Eq.~(\ref{ugdf}) 
is absorbed from the hard subprocess part describing the coupling of 
a hard gluon to the $q{\bar q}$ dipole, and gives us a hint that the
off-diagonal UGDF should be proportional to $\sim\sqrt{x_Pg(x_P)}$.

It is known that at some soft scale
$\mu_{\text{soft}}\sim \Lambda_\text{QCD}$ collinear PDFs like GRV
\cite{Gluck:1994uf} saturate at small $x'\ll x_P$, so one can introduce a function 
$\bar{R}_g(x',\mu_{\text{soft}}^2)$,
which is assumed to be slowly dependent on $x'$ in the case $x'\ll
x_P$:
 \begin{eqnarray}
 \sqrt{x_P}{\cal F}^{\text{off}}_g \simeq \bar{R}_g(x',\mu_{\text{soft}}^2)
 \sqrt{x_Pg(x_P,\mu_F^2)}\,
 f_{G}(\Delta_{\perp}^2).
 \label{ugdfsat}
 \end{eqnarray}
This factor $\bar{R}_g(x')$ contains all the soft
physics related with soft gluon couplings to the proton. It is interpreted as the 
square root of the gluon density at very small
$x'\ll x_P$ and soft scale $\mu_{\text{soft}}^2$. 
This is a non-perturbative object, which contributes to the overall
normalization and can be determined from data. As will be seen below, the
prescription (\ref{ugdfsat}) is consistent with the HERA data for
all available $M_X^2$ and $Q^2$.

There is a debate about what the power of the gluon density 
in the cross section should be (see \cite{BEHI05} and references therein). When squaring the amplitude containing (\ref{ugdfsat}), 
this model leads to a \textit{linear} dependence of the diffractive cross section on 
the gluon PDF. This linear dependence is the same as in the SCI model, 
where it describes both diffractive and non-diffractive events, 
and provides a continuous transition between the two types of events. 

This is in contrast to the \textit{quadratic} dependence on the gluon density often encountered in two-gluon 
exchange calculations of DDIS~\cite{WM99}. This arises from another prescription for the off-diagonal UGDF in the asymmetric limit $x'\ll x_P$ and $\mu_F^2\gg \Delta_{\perp}^2$, which in terms 
of the diagonal UGDF reads \cite{ugdfs}
 \begin{equation}
 f^{\mathrm{off}}_g(x_P,x',\Delta_{\perp}^2,\Delta_{\perp}^2,\mu_F^2)\simeq
 R_g(x')\; f_g(x_P,\Delta_{\perp}^2,\mu_F^2),
 \label{rg}
 \end{equation}
where the skewedness parameter\footnote{Our function $\bar{R}_g(x')$ is an analog of this skewedness factor.} $R_g\simeq 1.2-1.3$ is roughly
constant at HERA energies, and gives only a small
contribution to an overall normalization uncertainty. 
The factor $R_g$ can be approximately taken into account in this 
case by rescaling the $x_P$ argument in the diagonal UGDF as~\cite{Cudell:2008gv}
 \begin{equation}
 f^{\mathrm{off}}_g(x_P,x',\Delta_{\perp}^2,\Delta_{\perp}^2,\mu_F^2)\simeq
 f_g(0.41 \; x_P,\Delta_{\perp}^2,\mu_F^2)\,.
 \label{rg1}
 \end{equation}
Using the same Gaussian Ansatz for the intrinsic transverse momentum 
dependence as in Eq.~(\ref{ugdf}), we see that prescription (\ref{rg}) 
leads to a quadratic dependence of the diffractive 
structure function on the gluon PDF.

The unintegrated gluon density in the form (\ref{sqrt})
 reduces to the diagonal form (\ref{rg}) in the kinematical domain
where $x'\sim x_P$ and the hard gluon is soft, $\mu_F\sim
\mu_{\text{soft}}$. In this limit there is no QCD
factorization, so the hard and soft gluons must be taken into account 
together on equal footing. This may be the case at low $Q^2$ and $M_X^2$, when a 
larger contribution to the cross section
comes from relatively soft scales $\mu_F\lesssim1$ GeV, and Eq.~(\ref{ugdf})
reduces to
 \begin{eqnarray}
 \sqrt{x_P}{\cal F}^{\text{off}}_g \simeq
 0.5\,x_Pg(0.5\,x_P,\mu_F^2)\,
 f_{G}(\Delta_{\perp}^2).
 \label{ugdfrg}
 \end{eqnarray}
where the factor 0.5 appears from the equal momentum sharing between
active and screening gluons, since
$x_P$ is the sum of all gluon momentum fractions. In this sense, the
prescription (\ref{sqrt}) is more general since it describes both
$x'\ll x_P$ and $x'\sim x_P$ regimes, and contains
prescription (\ref{rg}) as a limiting case. 

Eq.~(\ref{ugdfrg}) also leads to a 
cross section with the gluon density in the second power. 
It is similar to the ``$R_g$-prescription'' (\ref{rg}) (if the Gaussian smearing like in
Eq.~(\ref{ugdf}) is adopted), and close to its phenomenological
form with rescaled argument (\ref{rg1}). This is more
conventional in the description of the exclusive processes \cite{WM99},
but is valid only for the symmetric case where
the two gluon exchanges carry longitudinal momentum fractions
close to each other, $x'\sim x_P$, and are connected to the same
factorization scale $\mu_F$. This case can also correspond to
the ``no-soft-exchange'' approximation, when the soft rescattering of the
on-shell partons in the final state is not taken into account (then $R_g=1$).

The above discussion leads to the conclusion that whether it is
appropriate to use prescription (\ref{ugdfsat}) or (\ref{ugdfrg}) 
depends on what kinematical regime one considers. The problem is
that $x'$ is not strictly constrained by kinematics. 
Its order of magnitude should be\footnote{This is similar to central exclusive production in
$pp$ collisions, where the screening gluons couple to
the triplet/antitriplet charges of the proton remnants, which have
predominantly equal momentum sharing $z\sim 1/2$,
and thus we have $x'\sim \Delta_{\perp}^2/s$.}
 \begin{eqnarray}
x'\sim \frac{\Delta_{\perp}^2}{z(1-z)W^2}\simeq
\frac{\mu_{\text{soft}}^2}{\mu_F^2}\, x_P \label{xpr}
 \end{eqnarray}
Thus the  $x'\ll x_P$ regime is realized in the
perturbative limit of large factorization scale $\mu_F$.
In the limit $\mu_F\to \mu_{\text{soft}}$, 
one instead has $x'\to x_P$, as naturally required by
matching the prescriptions of Eq.\ (\ref{ugdfsat}) and (\ref{ugdfrg}).

To summarize this section, we have formulated a framework for the gluon density needed as input to the calculation of the diffractive cross section. Since this describes soft QCD dynamics in the proton, there are necessarily some uncertainties. 
The precise HERA data on the diffractive cross section are, however, directly sensitive to this gluon density and may, therefore, be used to obtain new information about the gluon PDF at
extremely small $x'$ values and at different scales.

\section{Leading-order quark dipole contribution}

We are now fully equipped to derive the amplitude for the
dipole--proton interaction.

\subsection{Hard-soft factorization}

The total amplitude for $\gamma^*p\to Xp$ is decomposed
into longitudinal (L) and transverse (T) parts depending on the
photon polarization $\lambda_{\gamma}=0,\,\pm1$, and each part
can be written as a convolution of the hard and soft subprocess
amplitudes based on loop integration and cutting rules.

Starting from the general Sudakov decomposition of the total
screening gluon momentum $q_2=aq'+bP'+\Delta_{\perp}$ with $a,b\ll
x_P$, we can write down the amplitude, for example, for transversely
polarized photon as
\begin{align*}
&M^{\lambda_{\gamma}=\pm}_{\mp\pm}\sim\int\frac{d^2\Delta_{\perp}}{(2\pi)^2}\,
da db\\
&\times\,\delta\Big((k_1+k_2+q_2)^2-M_X^2\Big)\delta\Big((P'-q_2)^2\Big)\\
&\times\,M^{\text{hard}}(\Delta'_{\perp}) \;
M^{\text{soft}}(\Delta_{\perp}) \; (k_1^x\pm i k_1^y)\,.
\end{align*}
The $\delta$-function product can be rewritten as
\begin{eqnarray*}
\delta(\dots)\delta(\dots)\simeq\frac{1}{W^4}
\delta\Big(b-x_B\frac{\Delta_{\perp}^2}{Q^2}\Big)
\delta\Big(a+x_B\frac{\Delta_{\perp}^2}{Q^2}\Big),
\end{eqnarray*}
which takes care of the integrals over $a$ and $b$, leading to
\begin{eqnarray*}
M^{\lambda_{\gamma}=\pm}_{\pm\mp}=\frac{Q}{M_X}
\sqrt{\frac{2z}{1-z}}\,\int\frac{d^2\Delta_{\perp}}{(2\pi)^2} \
M^{\text{hard}}\, M^{\text{soft}}\\ \times (k_1^x\pm i k_1^y)\,,
\end{eqnarray*}
where, in the frame with $q'_{\perp}=0$,
\begin{eqnarray*}
k_1^x\pm i k_1^y=-(k_2^x\pm i k_2^y)\equiv k^x\pm i k^y.
\end{eqnarray*}
Analogously, the longitudinal contribution is
\begin{eqnarray*}
M^{\lambda_{\gamma}=0}_{\pm\mp}=
i\sqrt{z(1-z)}Q \int\frac{d^2\Delta_{\perp}}{(2\pi)^2} \,
M^{\text{hard}}\, M^{\text{soft}}.
\end{eqnarray*}

To calculate the Fourier transform of the total amplitude we use the
convolution theorem
\begin{eqnarray*}
h(q)=\int f(p)g(q-p)dp\quad\to\quad \hat{h}(x)=\hat{f}(x)\,
\hat{g}(x)
\end{eqnarray*}
where $\hat f$ denotes the Fourier transform of $f$. In our
amplitude this convolution is represented by the integral over
$\Delta_{\perp}$, while
$\delta=\sqrt{-t}=|\Delta'_{\perp}+\Delta_{\perp}|$ plays the role of
$q$. Thus, the inverse transformation over the impact parameter $b$ is
\begin{eqnarray*}
  M(\delta)\sim \int d^2b e^{-i{\bm
 \delta} {\bf b}} \hat{M}^\text{hard}({\bf b}) \, \hat{M}^\text{soft}({\bf b})\,.
\end{eqnarray*}
leading to factorization of the amplitude in $b$-space as a
direct consequence of $k_{\perp}$-factorization in impact parameter space.

\subsection{Hard part}

Consider first the hard gluon coupling to $q$ or $\bar q$ shown in
Fig.~\ref{fig:amp} in the rotated frame of reference with
$q_{\perp}=\Delta_{\perp}$ and $P_{\perp}=-\Delta_{\perp}$, where
$k'_{2\perp}=-k'_{1\perp}\equiv k'_{\perp}$. In what follows, we use
the relation for quark-gluon vertices in the eikonal approximation
\begin{equation*}
\bar{u}(k_2'+q_1)\, \gamma^\mu \, u(k_2') \simeq  2k_2^{\prime\mu},\qquad
q_1\ll k_2',
\end{equation*}
and for the product of the $t$-matrices in the large $N_c$ limit we have
\begin{eqnarray*}
t^a_{ij} t^a_{kl}\simeq T_F \delta_{ik}\delta_{jl}\,.
\end{eqnarray*}
The hard part,
describing the two possible couplings of the hard gluon to the
$q{\bar q}$ pair, can then be written as
 \begin{equation}
 M_{L,T}^\text{hard}(\Delta_{\perp},k'_{\perp})=\int d^2{\bf r}d^2{\bf b}\,
 \hat{M}_{L,T}^\text{hard}({\bf b},{\bf r}) e^{-i{\bf r}{\bf k}'_{\perp}}
 e^{-i{\bf b}{\bf \Delta}_{\perp}}\,, \label{Fourier-hard}
 \end{equation}
where $k'_{\perp}$ is the transverse momentum of a quark in the
intermediate state (see Fig.~\ref{fig:amp}), and the
Fourier-transformed hard amplitudes are given by
 \begin{align*}
 \hat{M}_{L}^\text{hard} =& \, i{\cal C} \, \alpha_s(\mu_F^2) \sqrt{\beta} \, W^3
 z^{3/2}(1-z)^{3/2}\,
 K_0(\varepsilon r){\cal V}({\bf b},{\bf r})\,,\\
 \hat{M}_{T,\pm}^\text{hard} =& \, i{\cal C} \alpha_s(\mu_F^2) \sqrt{\frac{2\beta}{1-\beta}}\,\frac{1}{\sqrt{x_P}} W^2
 z^{1/2}(1-z)^{3/2}\,
 \nonumber\\
\times&  \varepsilon K_1(\varepsilon r)  \frac{r_x\pm ir_y}{r}{\cal V}({\bf b},{\bf r})\,,
 \end{align*}
where ${\cal C}=8\pi e_q \sqrt{\pi \alpha_{em}}/N_c^2$, and $K_{0,1}$ are Bessel functions. 
The function ${\cal V}({\bf b},{\bf r})$ is the gluon density in impact parameter space defined as
 \begin{align} \nonumber
 {\cal V}({\bf b},{\bf r}) &=\frac{1}{\alpha_s(\mu_{\text{soft}}^2)}\,\int\frac{d^2\Delta_{\perp}}{(2\pi)^2}\,
 \sqrt{x_P}{\cal F}^{\text{off}}_g(x_P,\Delta_{\perp}^{2}) \\ \label{pdf}
 &\times
 \left\{e^{-i{\bf r}{\bf \Delta}_{\perp}}-e^{i{\bf r}{\bf
 \Delta}_{\perp}}\right\}\,
 e^{i{\bf b}{\bf \Delta}_{\perp}}\,.
 \end{align}
Here, ${\cal F}^{\text{off}}_g(x_P,\Delta_{\perp}^{2})$ is the
generalized UGDF, and $\mu_{\text{soft}}^2$ the typical soft scale of the process 
given by the gluon virtuality $\sim\Delta_{\perp}^2$. 
The factor containing $\alpha_s(\mu_{\text{soft}}^2)$ is introduced in the normalization of the soft 
part in order to compensate its absorption into the UGDF 
(see above). Inserting ${\cal F}^{\text{off}}_g$  from Eq.~(\ref{ugdfsat}), we finally get
 \begin{align}\nonumber
 {\cal V}({\bf b},{\bf
 r}) &= \frac{1}{\alpha_s(\mu_{\text{soft}}^2)}\,\frac{\bar{R}_g(x')}{(2\pi)^2}\,\sqrt{x_Pg(x_P,\mu_F^2)}\, \\ \label{Vdens}
&\times \left[e^{-\frac{\rho_0^2}{2}|{\bf b}-{\bf r}|^2}-
 e^{-\frac{\rho_0^2}{2}|{\bf b}+{\bf r}|^2}\right] .
 \end{align}
In the Fourier transformation we assumed slow evolution of
the QCD coupling $\alpha_s(\mu_{\text{soft}}^2)$, as is the case in the analytic perturbation theory discussed in next section or often assumed by freezing the coupling at very small $\mu_{\text{soft}}^2$. Thus, in the Gaussian model 
(\ref{ugdf}), the unintegrated gluon density in the impact parameter space 
${\cal V}({\bf b},{\bf r})$ is factorized into a collinear gluon density multiplied
by an ($r,b$)-dependent normal distribution.

\subsection{Soft part}\label{softpart}

We now turn to the soft subprocess amplitude, which can be
calculated order-by-order as follows. The softness of the
color-screening gluons with $x_i'\ll x_P$ implies that all
intermediate particles are on-shell, and that the
dipole size $r$ is not changed during the soft interactions. Cutting
the intermediate propagators we have only the phase shifts with the 
same origin as in Eq.~(\ref{Fourier-hard}), and a dependence on the soft momentum
exchanges $\Delta'_{i,\perp}$.

\begin{figure}[b]
 \centerline{\includegraphics[width=0.5\textwidth]{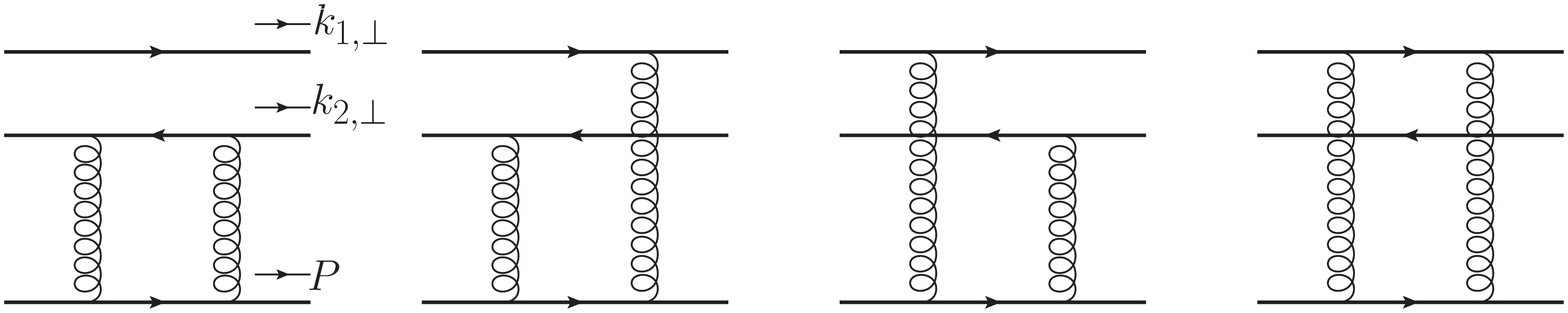}}
   \caption{\label{fig:2gresum}
   Illustration of resummation of the gluonic contributions to the
   (NLO) soft part $M_2^{\text{soft}}$ (two gluon exchanges) in the large $N_c$ limit.}
\end{figure}

In particular, for one and two soft gluon exchanges (in the large
$N_c$ limit) we obtain
 \begin{eqnarray*}
 e^{-i{\bf r}{\bf k}_{\perp}'}M^\text{soft}_{1}={\cal A}\,
 e^{-i{\bf r}{\bf k}_{\perp}} \frac{1}{{\Delta'}^{2}_{\perp}}
 \left[e^{-i{\bf r}{\bf \Delta}'_{\perp}}-1\right]\,,\\
 e^{-i{\bf r}{\bf k}_{\perp}'}M^\text{soft}_{2}=\frac{{\cal A}^2}{2!}\,
 e^{-i{\bf r}{\bf k}_{\perp}}\int\frac{d^2\Delta'_{2\perp}}{(2\pi)^2}
 \frac{1}{{\Delta'}^{2}_{1\perp}{\Delta'}^{2}_{2\perp}}\\
 \times
 \left[e^{-i{\bf r}{\bf \Delta}'_{\perp}}-e^{-i{\bf r}{\bf \Delta}'_{2\perp}}-
 e^{-i{\bf r}{\bf \Delta}'_{1\perp}}+1\right]\,,
 \end{eqnarray*}
where $\Delta'_{1\perp}=\Delta'_{\perp}-\Delta'_{2\perp}$ and ${\cal
A}=2\pi i\,C_F\, \alpha_s(\mu_{\text{soft}}^2)$ with $C_F\simeq T_F N_c$ in the
large $N_c$ limit. For example, the NLO gluonic contribution to the 
soft part $M_2^{\text{soft}}$ is represented by the four diagrams shown 
in Fig.~\ref{fig:2gresum}.

Fourier transformation with respect to ${\bf \Delta}'_{\perp}$ leads
to
 \begin{eqnarray} \label{resum}
 &&e^{-i{\bf r}{\bf k}_{\perp}'}\hat{M}^\text{soft}_{1}=
 e^{-i{\bf r}{\bf k}_{\perp}}\,{\cal A} \; {\cal W}({\bf b},{\bf r})\,,\\
 &&e^{-i{\bf r}{\bf k}_{\perp}'}\hat{M}^\text{soft}_{2}=
 e^{-i{\bf r}{\bf k}_{\perp}}\,\frac{{\cal A}^2\;
 {\cal W}({\bf b},{\bf r})^2}{2!}\,,\; \dots\nonumber
 \end{eqnarray}
where
 \begin{eqnarray}\label{Wq}
 {\cal W}({\bf b},{\bf r})=\frac{1}{2\pi}\ln\frac{|{\bf b}-{\bf r}|}
 {|{\bf b}|}\,.
 \end{eqnarray}
Continuing this procedure we see that summing over the
number of soft gluons in the final state leads to exponentiation in 
impact parameter space, so that for the total soft subprocess
amplitude we finally get
 \begin{eqnarray} \label{M-soft}
 e^{-i{\bf r}{\bf k}_{\perp}'}\hat{M}^\text{soft}({\bf b},{\bf r})=
 -e^{-i{\bf r}{\bf k}_{\perp}}\,(1-e^{{\cal A}
 \;{\cal W}({\bf b},{\bf r})})\,.
 \end{eqnarray}
A similar expression was previously derived in the case of scalar
Abelian gauge theory in Ref.~\cite{Brodsky}. Note that
$\hat{M}^\text{soft}({\bf b},{\bf r})$ is independent of the photon
polarization in the soft limit of small $\Delta'_{i,\perp}$.

As mentioned before, the soft gluon exchanges between the final
state partons occur at non-perturbatively small
longitudinal $x'$ and transverse momentum
transfer $\Delta'_{\perp}$  at some soft scale $\mu_{\text{soft}}$.
The strong coupling $\alpha_s(\mu_{\text{soft}}^2)$ is not
small in this case. There are several approaches for dealing with the Landau pole at
low momentum transfer (see e.g.\ Ref.~\cite{Brodsky:2010ur} and references therein).
We use the infrared-finite  Analytic Perturbation Theory (APT)~\cite{Shirkov:1997wi} approach  
to parametrize $\alpha_s(\mu_{\text{soft}}^2)$ at $\mu_{\text{soft}}^2\sim \Lambda_\text{QCD}^2$.
The analytic strong coupling $\alpha_s^{\text{APT}}(\mu^2)$ is stable
with respect to the choice of the QCD renormalization scheme,
higher-order radiative corrections and variations in $\Lambda_\text{QCD}$.
APT has also proved to give a
quantitative description of light quarkonium spectra within the
Bethe--Salpeter approach \cite{Baldicchi:2007ic} and DIS spin sum rules at low
$Q^2$ \cite{Pasechnik:2009yc}. 

In the one-loop case, the APT Euclidean function $\acal_1$, i.e.\ the
analyticized first power of the coupling $\alpha_s$ in the Euclidean
domain, is \cite{Shirkov:1997wi}
\begin{equation}
 \acal_1^{(1)}(\Delta_{\perp}^2)=\frac{1}{\beta_0}\left[\frac{1}{L}+
 \frac{\Lambda_\text{QCD}^2}{\Lambda_\text{QCD}^2-\Delta_{\perp}^2}\right],
 \;
 L=\ln\left(\frac{\Delta_{\perp}^2}{\Lambda_\text{QCD}^2}\right)
\end{equation}
where $\beta_0$ is the first coefficient of the QCD
$\beta$-function.

Since a significant contribution comes from the phase space region
with strongly uneven longitudinal momentum distribution between
the quark and the antiquark, and where $k_{\perp}$ is not very large, the
diffractive structure function becomes sensitive to the
model of the strong coupling used to calculate 
$\alpha_s^{\text{soft}}$, and hence to the typical soft
scale $\mu_{\text{soft}}$ of the process. To avoid this problem in
practice, in the soft regime we are considering, where 
$\Delta_{\perp}\sim \Lambda_\text{QCD}$, we
do not extract the value of $\mu_{\text{soft}}$ but rather fix the
coupling at $\alpha_s^{\text{soft}}=\acal_1(\Lambda_\text{QCD})\simeq 0.7$.

Having all the L and T amplitudes, we can now finally rotate our
frame of reference in the transverse plane to one with
$q'_{\perp}=0$ and $P'_{\perp}=0$, where $k_{\perp}$ is defined as
in Eq.~(\ref{kt}). In this frame, the variables $b$ and $r$ are
the impact parameter and the $q{\bar q}$ dipole size in the final state,
respectively.

\subsection{Diffractive structure function}

Let us turn to the cross section. We are interested in the
case when the proton remnant forms a particle in the final state
with invariant mass close to the proton mass, so we have a three
particle phase space. We may write in
general
\begin{eqnarray*}
&&q'_0\frac{d^3\sigma^{\gamma^*p\to Xp'}}{d^3 {\bf
q}'}=\frac{1}{2\lambda^{1/2}(W^2,m_p^2,-Q^2)}\\
&&\times\,\int \frac{d^3{\bf k}_1}{(2\pi)^3\,2k_1^0}
\frac{d^3{\bf k}_2}{(2\pi)^3\,2k_2^0} \frac{d^3{\bf P}'}{(2\pi)^3\,2P'_0}\,
q'_0\delta^{(3)}({\bf q}'-{\bf k}_1-{\bf k}_2)\\
&&\times\,(2\pi)^4\delta^{(4)}(q+p-k_1-k_2-P')\sum_{\lambda_{q},\lambda_{\bar
q},\lambda_{\gamma}} |M^{\lambda_{\gamma}}_{\lambda_{q}\lambda_{\bar
q}}|^2\,.
\end{eqnarray*}
In the large $W$ limit the flux factor is
$2\lambda^{1/2}(W^2,m_p^2,-Q^2)\simeq 2W^2$. The left
hand side can be transformed to
\begin{eqnarray*}
q'_0\frac{d^3\sigma^{\gamma^*p\to Xp'}}{d^3 {\bf q}'}\simeq
\frac{W^2}{\pi}\frac{d^2\sigma^{\gamma^*p\to Xp'}}{dM_X^2dt}=
-\frac{W^2\beta^2}{\pi Q^2}\frac{d^2\sigma^{\gamma^*p\to
Xp'}}{d\beta dt}\,.
\end{eqnarray*}
The diffractive structure functions have simple relations to the
corresponding differential cross sections
\begin{eqnarray*}
&&x_P F_{L,T}^{D(4)}(\beta, x_P, Q^2, t)=-\frac{Q^2\beta}{4\pi^2\alpha_{em}}
\frac{d^2\sigma^{\gamma^*p\to Xp'}_{L,T}}{d\beta dt},\\
&&x_P F_{L,T}^{D(3)}(\beta, x_P,
Q^2)\simeq-\frac{Q^2\beta}{4\pi^2\alpha_{em}}\frac{1}{B_D}\frac{d^2\sigma^{\gamma^*p\to
Xp'}_{L,T}}{d\beta dt}\Big|_{t=0},
\end{eqnarray*}
assuming an exponential $t$-dependence $\sim\exp({B_Dt})$ of
the cross section on the diffractive slope $B_D$. The
$\delta$-functions remove the integrals over ${\bf P}'$ and one of
the quark momenta, say, ${\bf k}_1$, and we get
 \begin{align*}
&\frac{d^2\sigma^{\gamma^*p\to Xp'}}{d\beta dt}\simeq
-\frac{Q^2}{4\beta^2}\frac{\pi}{(2\pi)^5}
\frac{1}{W^6}\\
\times\,\int\frac{d^3{\bf
k}_2}{z(1-z)}\;&\delta(q_0+P_0-k_1^0-k_2^0-P'_0)\sum_{\lambda_{q},\lambda_{\bar
q},\lambda_{\gamma}} |M^{\lambda_{\gamma}}_{\lambda_{q}\lambda_{\bar
q}}|^2
 \end{align*}
The remaining $\delta$-function removes the integral over
$k_{2z}\simeq zW/2$. The last phase space integration is rather
trivial,
\begin{eqnarray*}
\int d^2{\bf
k}_{\perp}=2\pi\,\frac12\int_0^{k_{\perp,\text{max}}^2}
dk_{\perp}^2=\pi M_X^2\int_0^{\frac12}dz (1-2z)\,.
\end{eqnarray*}
Finally, in the full phase space we have to take into account an extra
factor of two due to the symmetry with respect to the interchange
$z\leftrightarrow 1-z$.

Straightforward calculation leads to the following expressions for
the longitudinal and transverse fully-unintegrated diffractive
structure functions $F^{D,(4)}_{L,T}(x_P,Q^2,\beta,t)$:
  \begin{align}
x_PF_L^{D(4)} &= {\cal S}\,Q^4M_X^2
\int_{z_{min}}^{\frac12}dz (1-2z)\, z^2(1-z)^2 |J_L|^2 \label{FL}\\
x_PF_T^{D(4)} &= 2{\cal S}\, Q^4\int_{z_{min}}^{\frac12}dz (1-2z) \left\{(1-z)^2+z^2\right\} |J_T|^2 \label{FT},
 \end{align}
where ${\cal S}={\sum_q e_q^2}/({2\pi^2N_c^3})$ sums over light quark charges $e_q$, and
 \begin{align} \nonumber
 J_L=i\alpha_s(\mu_F^2)
 \int d^2{\bf r}d^2{\bf b}\, e^{-i{\bm \delta} {\bf b}}
 e^{-i{\bf r}{\bf k}_{\perp}}\,K_0(\varepsilon r) \nonumber \\
 \times\,{\cal V}({\bf b},{\bf r})
 \Big[1-e^{{\cal A}{\cal W}}\Big], \label{JL} \\
 J_T=i\alpha_s(\mu_F^2)
 \int d^2{\bf r}d^2{\bf b}\, e^{-i{\bm \delta}
 {\bf b}}e^{-i{\bf r}{\bf k}_{\perp}}\,\varepsilon K_1(\varepsilon r)\nonumber \\
 \times\,\frac{r_x\pm ir_y}{r} \, {\cal V}({\bf b},{\bf r})
 \Big[1-e^{{\cal A}{\cal W}}\Big]\,. \label{JT}
 \end{align}
These are the general expressions of the
QCD-based soft multiple gluon rescattering model.

\subsection{Physical interpretation and simplification}\label{interpretation}

We have now derived Eqs.\ (\ref{FL}--\ref{JT}), which describe the 
diffractive structure function. These 
have non-perturbative soft gluon exchanges as important ingredients, and 
to calculate these exchanges we have had to make some model 
assumptions. Some of these assumptions have already been discussed above:\ 
we treat the coupling to the quarks using the strong coupling obtained in 
APT and the coupling to the proton remnant using the function $\bar R_g$. 
Moreover, we extrapolate perturbation theory and assume a perturbative 
propagator for the gluons. The infrared logarithmic divergences in these 
gluon propagators, which appear at each order in the resummation 
(see Eqs.\ (\ref{resum},\ref{Wq})), disappear
when the gluon exchanges are resummed to all orders (Eq.\ (\ref{M-soft})).

There is one additional model assumption, as we will explain shortly, but let 
us first discuss a physical argument based on effective field theory principles, 
or equivalently, on the uncertainty principle: A gluon with momentum $k$ has a 
``resolution power,'' or minimal scale of an object it can resolve, of order 
$1/k$. Put another way, physics should not depend on scales much smaller 
than the resolution scale.

The hard gluon in our calculation can resolve the $q{\bar q}$ dipole with 
transverse size $r$, allowing us to apply perturbation theory to the hard part. 
On the other hand, the soft gluons do not carry significant longitudinal 
momentum fractions, and only small transverse momenta 
$\Delta_{\perp}\sim \delta=\sqrt{-t}$, so their ``maximal resolution'' 
scale is  $b\sim 1/\delta$.
This means that the screening gluons cannot dynamically resolve the internal 
structure of a small $q{\bar q}$ dipole with size $r\ll b$. However, in 
constructing our model, we extrapolate perturbative QCD to the non-perturbative 
regime and assume that the soft gluons couple individually to the quark and antiquark, 
since the essential point of the dynamics here is the color exchange and not the 
momentum transfer. This is in a similar vein to using quark currents in hadronic 
matrix elements, such as form factors. The underlying quark and gluon dynamics 
is still important even at very low scales (see e.g.\ Ref.~\cite{Brodsky:2010ur} for a discussion of this).
In this way there is a continuous transition between soft 
perturbative and soft non-perturbative gluons. 

As they stand, the integrals in Eqs.\ (\ref{JL},\ref{JT}) exhibit unphysical 
singularities in the angular integrations. This, however, is because 
of our model assumption, which so far does not fully take into account the resolution 
power argument. Since the gluons are soft, physics should not depend on the 
orientation of the $q{\bar q}$ dipole with respect to the impact parameter. 
This will regulate the unphysical singularities in the angular integration in 
Eqs.~(\ref{JL},\ref{JT}). This will also allow us to evaluate the integrals 
analytically, and we will use this in our calculations below. We argue that the 
resulting expressions, Eqs.\ (\ref{JL2},\ref{JT2}) below, are the physically correct 
expressions for $J_L$ and $J_T$ to use in Eqs.\ (\ref{FL},\ref{FT}).

The expression (\ref{Vdens}) can be considered as a
model for the unintegrated gluon density in impact parameter
space. In particular, it defines the probability to probe a gluon at
 impact distance $b$ from the proton center with momentum
$\Delta_{\perp}\sim 1/b$ by a hard $q\bar{q}$ dipole with small
size $r\ll b$, where the quarks carry the hard momentum $k_{\perp}\sim
1/r$. The process is considered at a factorization scale
equal to the quark virtuality $\mu_F^2$. The gluons
cannot resolve scales below the dipole size $r$. Therefore, the gluon density cannot depend on the
orientation of the $q\bar{q}$ dipole with respect to ${\bf b}$, i.e., on the
angle between ${\bf r}$ and ${\bf b}$. Also, in this
Gaussian model there is no physical reason for an asymmetry of the
UGDF with respect to the direction of the vector ${\bf b}$. Thus we rewrite our
expression (\ref{Vdens}) in the following way
\begin{equation}\label{Vdensmod}
{\cal V}\simeq \frac{1}{\alpha_s(\mu_{\text{soft}}^2)}\,\frac{\bar{R}_g(x')}{2\pi^2}\sqrt{x_Pg}
\,e^{-\frac{\rho_0^2}{2}(b^2+r^2)}\sinh(\rho_0^2\,br).
\end{equation}
In the small dipole limit $r\ll b$ this becomes
 \begin{eqnarray}
{\cal V}(b,r)\simeq \frac{1}{\alpha_s(\mu_{\text{soft}}^2)}\,\bar{R}_g(x')\frac{\rho_0^2}{2\pi^2}\,
\sqrt{x_Pg}\,br\,e^{-\frac{\rho_0^2}{2}b^2},\label{Vbr}
\end{eqnarray}
which will be used below to obtain the formula for the DDIS amplitudes.

We can check our formalism by taking the small coupling limit
$\alpha_s(\mu_\text{soft}^2)\sim\alpha_s(\mu_F^2)\ll1$, where we can approximate
 \begin{eqnarray}
1-e^{{\cal A}{\cal W}}\simeq
-i\alpha_s(\mu_\text{soft}^2)C_F\frac{r}{b}\,.\label{altwogl}
 \end{eqnarray}
In the longitudinally polarized case, the Fourier integrals are then
reduced to Hankel transforms, leading to
 \begin{eqnarray}\nonumber
 J_L\simeq 8\bar{R}_g(x') \,\alpha_s(\mu_F^2)C_F
 \sqrt{x_Pg}\,e^{\frac{t}{2\rho_0^2}}
 \frac{\varepsilon^2-k_{\perp}^2}{(\varepsilon^2+k_{\perp}^2)^3}.
 \end{eqnarray}
Thus, in the limit $\alpha_s\ll1$ our model successfully
reproduces the standard leading-order two-gluon amplitude \cite{gwus}
and leads to the correct exponential $t$-dependence of the
cross-section $\sim\exp( B_Dt)$ with diffractive slope
$B_D\equiv1/\rho_0^2=6.9\pm 0.2$ GeV$^{-2}$ known from HERA data
\cite{Chekanov:2008fh}. This gives $\rho_0\simeq 380$ MeV,  close to
the value of $\Lambda_\text{QCD}$. Thus, the Gaussian 
width $\rho_0$ physically corresponds to the effective 
transverse size of the proton.

However, the strong coupling $\alpha_s(\mu_{\text{soft}}^2)$ is not
small in the case of small momentum transfers $\lesssim
\Lambda_\text{QCD}$, and we cannot calculate the integral in $J_L$ in
general form analytically. The soft phase ${\cal A}{\cal W}$ is not in general 
small in the Fourier transformation, and in evaluating
the Fourier integrals in Eqs.~(\ref{JL},\ref{JT}) we should not
impose the ${\cal W}\ll1$ condition, but rather keep the exponent
$\exp({\cal A}{\cal W})$ with imaginary ${\cal A}$. This produces an
extra phase shift in the Fourier transform over ${\bf r}$, coming
from the soft gluon exponentiation in the large $N_c$ limit.
Employing the ``maximum resolution'' argument introduced above, we
can write
\begin{eqnarray}\label{W}
e^{{\cal A}{\cal W}({\bf b},{\bf r})}\simeq e^{-i {\bf r}{\bm
\eta}}\,, \quad {\bm \eta}=\alpha_s(\mu_\text{soft}^2)C_F\frac{{\bf
b}}{b^2} 
\end{eqnarray}
In the longitudinally polarized case the result of the Fourier
integration over ${\bf r}$ is the Hankel transformation of
$K_0(\varepsilon r)\,r$ with respect to the momenta ${\bf
k}_{\perp}$ and ${\bf k}_{\perp}+{\bm \eta}$. We obtain
\begin{align} \nonumber
&J_L\simeq
\bar{R}_g \, \frac{\alpha_s(\mu_F^2)}{\alpha_s(\mu_\text{soft}^2)}\,
\sqrt{x_Pg}\,\frac{\rho_0^2}{\pi}\int d^2{\bf b}\,
e^{-i{\bm \delta}
{\bf b}}\,e^{-\frac12\rho_0^2b^2}\\
&\times\,b\Bigg[\frac{2\varepsilon^2E\Big(-\frac{k_{\perp}^2}{\varepsilon^2}\Big)
-\left(\varepsilon^2+k_{\perp}^2\right)
K\Big(-\frac{k_{\perp}^2}{\varepsilon^2}\Big)}{\varepsilon
\left(\varepsilon^2+k_{\perp}^2\right)^2}- \label{J_L}\\
&\frac{2\varepsilon^2E\Big(-\frac{(k_{\perp}+\eta)^2}{\varepsilon^2}\Big)-
\left(\varepsilon^2+(k_{\perp}+\eta)^2\right)
K\Big(-\frac{(k_{\perp}+\eta)^2}{\varepsilon^2}\Big)}{\varepsilon
\left(\varepsilon^2+(k_{\perp}+\eta)^2\right)^2}\Bigg] \nonumber
\end{align}
in terms of the complete elliptic integrals of the first and second
kind, $K(x)$ and $E(x)$, respectively. In the forward limit of small
$\delta\ll k_{\perp}$ we expect $k_{\perp}\gg \eta$,
$\eta\equiv|{\bm \eta}|$. 

There is a similar simplification
in momentum space for the hard momentum $k_{\perp}\sim 1/r$, i.e.,
${\bf k}_{\perp}{\bm \eta}\sim k_{\perp}\eta$, neglecting the dependence 
on the direction of the $q\bar q$
transverse momentum in the isotropic color field of the proton remnant.

Further, we expand the integrand in Eq.~(\ref{J_L}) in
$\xi=k_{\perp}\eta+\eta^2\ll k_{\perp}^2$, and
keep only the leading term in $\xi$. Taking the last Fourier
integral gives
\begin{equation}\nonumber
\int d^2{\bf b}\,e^{-i{\bm \delta}{\bf
b}}\,e^{-\frac{\rho_0^2}{2}b^2}
\bigg(k_{\perp}+\frac{v}{b}\bigg)=\frac{2\pi}{\rho_0^2}\,{\cal U}(t),
\end{equation}
where
\begin{equation}
{\cal U}(t)=k_{\perp}\,e^{\frac{t}{2\rho_0^2}}+\sqrt{\frac{\pi}{2}}\,
v\,\rho_0\,e^{\frac{t}{4\rho_0^2}}\,I_0\bigg(\frac{-t}{4\rho_0^2}\bigg),
\label{bint}
\end{equation}
$I_0$ is a modified Bessel function, and $v=\alpha_s(\mu_\text{soft}^2)C_F$. The second term is an NLO
contribution since it is proportional to the $\alpha_s$ in $v$,
and typically in the forward limit $t\ll 1\,\mathrm{GeV}^2$ and in
the hard momentum transfer limit it is much smaller than the leading
term (we do not consider large $t$, where the whole formalism here does not apply). However,
this term is the only leading term which survives in the limit when
both $k_{\perp}\to 0$ and $|t|\sim 1/b^2\to 0$, so we have
to take it into account.

As regards the $t$-dependence, the second term in Eq.~(\ref{bint}) decreases at
large $t$, but not as rapidly as the first term. To
good approximation, the integral over $t$ in the cross
section can be written as
\begin{eqnarray}\nonumber
\int_0^1 dt\,{\cal U}(t)^2\simeq \rho_0^2(k_{\perp}+\sigma_0
v)^2,\quad \sigma_0=0.73\,\mathrm{GeV}.
\end{eqnarray}
The second term must be taken into account when $k_{\perp}$ is $\lesssim 1$~GeV.

Straightforward calculation leads to the following result for the
longitudinal contribution,
\begin{align}
&J_L\simeq  \label{JL2}
\bar{R}_g\sqrt{x_Pg}\,\frac{\alpha_s(\mu_F^2)C_F\,{\cal
U}(t)}{2\varepsilon^3\left(k_{\perp}^2+\varepsilon^2\right)^3}
\Bigg[\left(k_{\perp}^2+\varepsilon ^2\right)\\
&\times\Bigg\{8
K\left(-\frac{k_{\perp}^2}{\varepsilon^2}\right)\varepsilon^2+ \nonumber
\pi\Bigg(2\,_2F_1\left(\frac{1}{2},\frac{3}{2};2;-\frac{k_{\perp}^2}{\varepsilon^2}\right)
\varepsilon^2 \\
&+\left(k_{\perp}^2+\varepsilon^2\right)\,_2F_1\left(\frac{3}{2},\frac{3}{2};2;
-\frac{k_{\perp}^2}{\varepsilon^2}\right)\Bigg)\Bigg\}-32\varepsilon^4
E\left(-\frac{k_{\perp}^2}{\varepsilon^2}\right)\Bigg], \nonumber
\end{align}
and for the transverse contribution,
\begin{align}
&J_T\simeq \label{JT2}
\bar{R}_g\sqrt{x_Pg}\,\frac{\alpha_s(\mu_F^2)C_F\,{\cal
U}(t)}{2\varepsilon k_{\perp}^3
\left(k_{\perp}^2+\varepsilon^2\right)^3}\\  \nonumber
&\times\Bigg[8\left(\varepsilon^6+3k_{\perp}^2\varepsilon^4-
2k_{\perp}^4\varepsilon^2\right)E\left(-\frac{k_{\perp}^2}{\varepsilon^2}\right)-
\left(\varepsilon^2+k_{\perp}^2\right)\\  \nonumber
&\qquad
\times\Bigg\{\pi\Bigg(\left(\varepsilon^2-k_{\perp}^2\right)\,
_2F_1\left(\frac{1}{2},\frac{3}{2};2;-\frac{k_{\perp}^2}{\varepsilon^2}\right)\\  \nonumber
&\qquad
+\left(\varepsilon^2+k_{\perp}^2\right)\,_2F_1\left(\frac{3}{2},\frac{3}{2};2;-\frac{k_{\perp}^2}
{\varepsilon^2}\right)\Bigg)k_{\perp}^2\\  \nonumber
&\qquad +8\left(\varepsilon^4+2 k_{\perp}^2 \varepsilon^2\right)
   K\left(-\frac{k_{\perp}^2}{\varepsilon^2}\right)\Bigg\}\Bigg],
\end{align}
where $_2F_1$ is the hypergeometric function and $E$ and $K$ are the complete elliptic integrals as above.

\section{Gluon contribution to the diffractive structure function}

In the large-$M_X$ limit, gluon emission may be important.
In principle, gluons may be radiated from both the
$q{\bar q}$ dipole and the hard gluon. The gluons
emitted from the quarks are dominantly soft and move collinearly with the
quarks, and do not significantly change the invariant mass of the final system
$X$. Rather, they dress the quarks to build up
their effective mass $m_q^{\text{eff}}$, which is, in general, a
function of the two hard scales $Q^2$ and $M_X^2$. This
mass parameter may be treated as a
constituent quark mass. In the current work we do not make
 predictions for $m_q^{\text{eff}}$, but instead extract it
from  data.

The scale dependence of the effective quark mass in processes
with two hard scales like the one under consideration may be
complicated. This will be discussed in connection with the 
numerical results in Sec.~\ref{numresults}.

\subsection{Kinematics}

\begin{figure}[tb]
 \centerline{\includegraphics[width=0.25\textwidth]{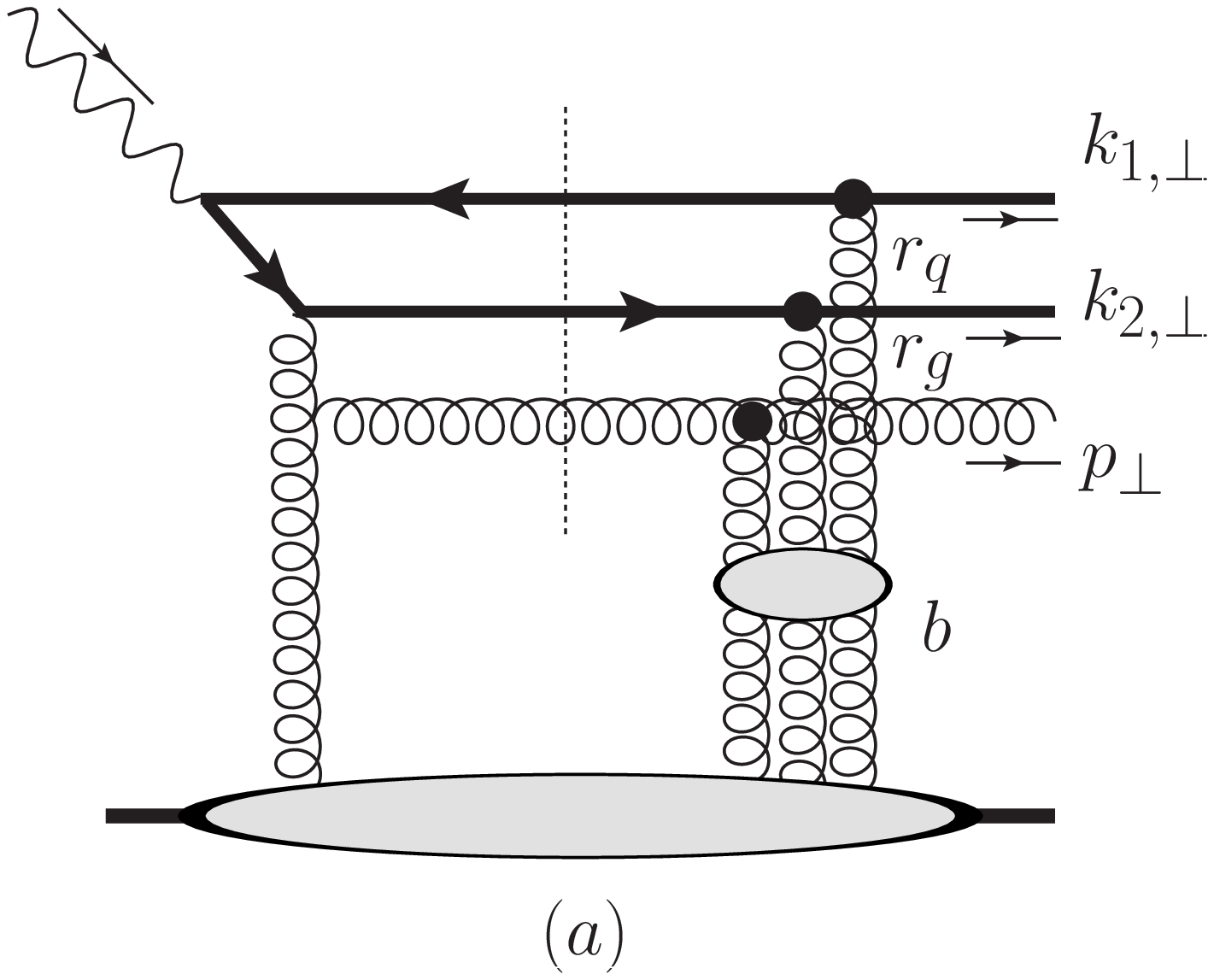}
\includegraphics[width=0.25\textwidth]{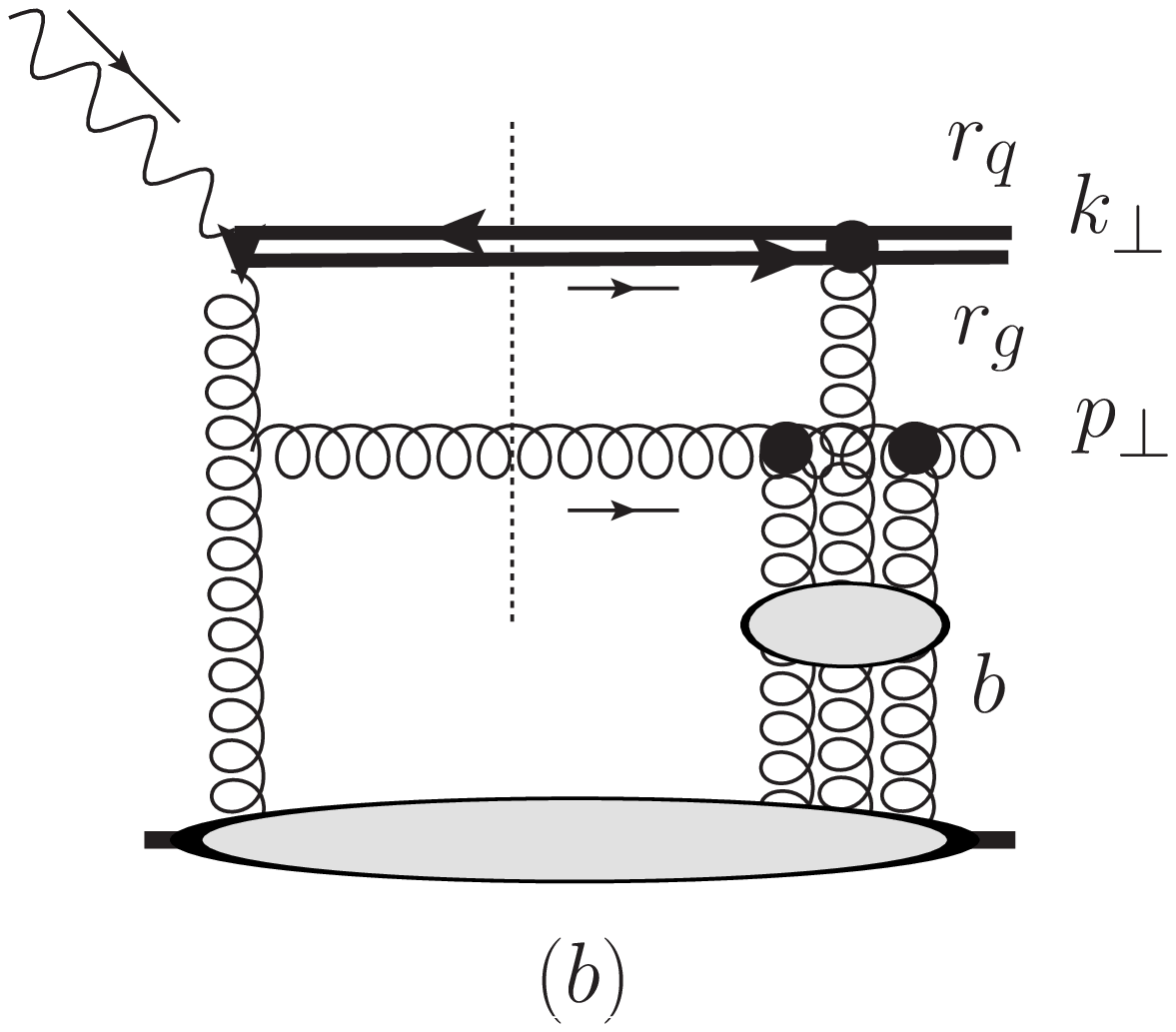}}
   \caption{\label{fig:qqg}
   Typical diagram for gluon emission in the
   DDIS final state (a), and its asymptotic limit reducing
   to the ``gluonic dipole'' contribution
   for $\beta\to0$ and $p_{\perp}\ll k_{\perp}$ (b).}
\end{figure}

The small $\beta\to0$ limit is, therefore, driven by gluon emission from 
the hard gluon, as illustrated in Fig.~\ref{fig:qqg}.
The kinematics of the process in the $XP'$ c.m.s.\ frame, where ${\bf
k}_{1,\perp}+{\bf k}_{2,\perp}=-{\bf p}_{\perp}$, is given by the
Sudakov decompositions
\begin{align} \nonumber
&k_1=(1-z-z')q'+n_1P'+k_{1,\perp},\\
&k_2=zq'+n_2P'+k_{2,\perp},\label{sudqqg}\\
&p=z'q'+n_3P'+p_{\perp}\,. \nonumber
\end{align}
with $n_1+n_2+n_3=0$. Analogously to the $q{\bar
q}$ case, we obtain the following expression for the invariant
mass $M_X$ in terms of momentum fractions $z,\,z'$:
\begin{align} \label{MXg1}
M_X^2=\frac{z
k_{1,\perp}^2+(1-z-z')k_{2,\perp}^2+(1-z')m_q^2}{z(1-z-z')}+\frac{p_{\perp}^2}{z'}.
\end{align}
$M_X$ can also be represented in terms of the invariant
mass of the $q{\bar q}$ system as
\begin{eqnarray} \label{MXg2}
M_X^2=\frac{M_{q{\bar
q}}^2}{1-z'}+\frac{p_{\perp}^2}{z'(1-z')},\quad
(k_1+k_2)^2=M_{q{\bar q}}^2\,.
\end{eqnarray}

\subsection{Soft gluon resummation and the gluonic dipole limit}

The $q{\bar q}g$-system scatters off the proton by exchanging
soft gluons, in the same way as the $q{\bar q}$-system above, and also
here the gluon exchanges can be resummed. We have two independent
transverse momenta, $k'_{1,\perp}$ and $k'_{2,\perp}$, outgoing 
from the hard subprocess, corresponding to impact
parameters $r_1$ and $r_2$. Proceeding as for the $q \bar q$ case, 
we obtain for the $q{\bar q}g$ case
 \begin{eqnarray*}
 e^{-i{\bf r}_1{\bf k}_{1,\perp}'}e^{-i{\bf r}_2{\bf k}_{2,\perp}'}
 M^\text{soft}_{1,q{\bar q}g}=
 e^{-i{\bf r}_1{\bf k}_{1,\perp}}e^{-i{\bf r}_2{\bf k}_{2,\perp}}
 \frac{1}{{\Delta'}^{2}_{\perp}}\\
 \times\left[{\cal A}e^{-i{\bf r}_1{\bf \Delta}'_{\perp}}+
 {\cal A}e^{-i{\bf r}_2{\bf \Delta}'_{\perp}}+{\cal A}_g\right]\,,
 \end{eqnarray*}
where the prefactor ${\cal A}$ was
introduced above for the gluon coupling to a $q\bar q$ dipole, and
${\cal A}_g$ corresponds to the case with a gluon coupling to a gluon
in the $q{\bar q}g$-system. By explicit calculation of the color
factors it can be shown that in the large-$N_c$ limit ${\cal A}_g=-2{\cal A}$. This
allows us to perform the Fourier transformation of the soft
part over ${\Delta'}_{\perp}$ for any number of exchanged gluons and
to resum them in the same way, as for $q{\bar q}$
dipole rescattering. For one and two gluon exchanges we have
 \begin{align*}
 e^{-i{\bf r}_1{\bf k}_{1,\perp}'}e^{-i{\bf r}_2{\bf k}_{2,\perp}'}
 &\hat{M}^\text{soft}_{1,q{\bar q}g}=
 e^{-i{\bf r}_1{\bf k}_{1,\perp}}e^{-i{\bf r}_2{\bf k}_{2,\perp}}\\
 &\times{\cal A}\;\Big[{\cal W}({\bf b},{\bf r}_1)+
 {\cal W}({\bf b},{\bf r}_2)\Big]\,,\\
 e^{-i{\bf r}_1{\bf k}_{1,\perp}'}e^{-i{\bf r}_2{\bf k}_{2,\perp}'}
 &\hat{M}^\text{soft}_{2,q{\bar q}g}=
 e^{-i{\bf r}_1{\bf k}_{1,\perp}}e^{-i{\bf r}_2{\bf k}_{2,\perp}}\\
 &\times\frac{{\cal A}^2}{2!}\;
 \Big[{\cal W}({\bf b},{\bf r}_1)+{\cal W}({\bf b},{\bf r}_2)\Big]^2\,,\;
 \nonumber
 \end{align*}
where ${\cal W}({\bf b},{\bf r})$ is defined above in
Eq.~(\ref{Wq}). Summing over the number of soft gluons in the final
state leads to exponentiation in impact parameter space, i.e.,
 \begin{align} \label{M-soft-qqg}
 &e^{-i{\bf r}_1{\bf k}_{1,\perp}'}e^{-i{\bf r}_2{\bf k}_{2,\perp}'}
 \hat{M}^\text{soft}({\bf b},{\bf r}_1,{\bf r}_2)=\\
 &-e^{-i{\bf r}_1{\bf k}_{1,\perp}}e^{-i{\bf r}_2{\bf k}_{2,\perp}}\,
 \Big[1-e^{{\cal A}
 [{\cal W}({\bf b},{\bf r}_1)+{\cal W}({\bf b},{\bf r}_2)]}\Big]\,.
 \nonumber
 \end{align}

Let us now focus on the leading asymptotic behavior of the diagram 
in Fig.~\ref{fig:qqg}(a) in the limit
$\beta\to0$. In this limit the hard scale of the process
$\mu_F^2\sim Q^2/\beta$ becomes very large.
From Eqs.~(\ref{MXg1}) or (\ref{MXg2}) we see that the
$M_X^2\to\infty$ limit is realized when $z'\ll z$ (more precisely
$z'\to0$), so the invariant mass of the $q{\bar q}g\equiv X$ system is
\begin{equation} \label{MXg3}
M_X^2\simeq M_{q{\bar q}}^2+\frac{p_{\perp}^2}{z'}\gg M_{q{\bar
q}}^2,
\end{equation}
where
\begin{equation}
M_{q{\bar q}}^2\simeq
\frac{k_{\perp}^2+m_q^2}{z(1-z)}\ll M_X^2.
\end{equation}

Consider first the limit where the gluon transverse momentum $p_{\perp}$ 
is small, such that $|{\bf k}_{1,\perp}|\simeq|{\bf k}_{2,\perp}|\gg 
|{\bf p}_{\perp}|$. In impact parameter space this kinematical configuration
corresponds to the diagram shown in Fig.~\ref{fig:qqg}(b). 
In this limit the $q\bar q$ pair is very small, i.e., we have strong 
ordering in impact parameter space, which
can be written as $r_q\ll r_g\ll b$. In  color space the
$q\bar q$ pair can be considered as a single gluon, and
we  consider ``gluonic dipole'' scattering off the target. This
is consistent with our expression for the corresponding soft
part (\ref{M-soft-qqg}), which in the limit $r_1\simeq
r_2\equiv r_g$ reduces to $\sim 1-\exp(2{\cal A}{\cal W}({\bf
b},{\bf r}_g))=1-\exp(-{\cal A}_g{\cal W}({\bf b},{\bf r}_g))$,
corresponding to the amplitude for soft gluon--gluon scattering. 
This reproduces the conventional $gg$ dipole result \cite{gwus} in the
small $r_g\ll b$ limit, in which the amplitude of the gluonic dipole
scattering differs by a factor of $N_c/C_F\simeq 1/T_F=2$ from
the amplitude of the $q{\bar q}$ scattering. Indeed, from our model
it follows that in this limit
\begin{align}\nonumber
A^{\text{soft}}_{gg}&=1-\exp(-{\cal A}_g{\cal W}({\bf b},{\bf
r}_g))  \\
&\simeq 2i\alpha^{\text{soft}}_sC_F\frac{r}{b}=-2
A^{\text{soft}}_{q{\bar q}} \label{ggqq}
\end{align}
as compared to Eq.~(\ref{altwogl}).

However, this limiting case cannot give a leading contribution 
to the diffractive structure function at large $M_X$ because of 
the smallness of the transverse 
momentum of the final state gluon $p_{\perp}\ll k_{1,2\perp}$. 
Due to Eq.~(\ref{MXg2}) the larger gluon $p_{\perp}$, the larger 
invariant mass $M_X$ is produced. At the same time, $p_{\perp}$ 
can not be significantly larger than the quark and antiquark transverse momenta 
$k_{1,2\perp}$. Due to momentum conservation, the maximal $M_X$ at fixed 
$z'$ occurs in the limit $p_{\perp}\sim k_{1,\perp}\gg k_{2,\perp}$, which 
corresponds to $r_1\ll r_2$ in impact parameter space, leading to  
${\cal W}({\bf b},{\bf r}_1)\ll{\cal W}({\bf b},{\bf r}_2)$. 
From Eq.~(\ref{M-soft-qqg}), this corresponds to the 
situation when only the $q\bar q$ component of the $q{\bar q}g$ 
system scatters off the target with soft part $A^{\text{soft}}_{q{\bar q}}$. 
This purely kinematical 
argument is compatible with an observation~\cite{Marquet:2007nf} 
with respect to models for parton saturation~\cite{satmod},
that the $q{\bar q}g$ and $q{\bar q}$ dipole contributions
should saturate to the same value, i.e. $A^{\text{soft}}_{q{\bar q}g}
\simeq A^{\text{soft}}_{q{\bar q}}$ at large invariant masses $M_X$. 
In particular, this means that the scattering of the $q{\bar q}g$ system
off the proton can not be reduced to the scattering of the $gg$ dipole.

\subsection{Leading $q{\bar q}g$ contribution to the diffractive
structure function}

We argued above that the leading $q{\bar q}g$ 
contribution to the diffractive structure function in the large
$M_X$ limit comes from on-shell gluon emission from the hard
gluon as in Fig.~\ref{fig:qqg}(a). It is clear from
Eq.~(\ref{MXg3}) that the relevant limit $M_X\to\infty$ corresponds
to essentially on-shell gluon emission with $z'\ll z$. 
The corresponding gluon propagator can be only slightly off-shell
to give a leading contribution to the cross section. In this case
the $q{\bar q}$ pair takes most of the longitudinal momentum
 of the $X$ system, and kinematically there is no symmetry
with respect to interchange $z'\leftrightarrow 1-z'$ in such a $q{\bar
q}g$ system, whereas for a $q{\bar q}$ dipole this symmetry
$z\leftrightarrow 1-z$ holds explicitly. If one allows
the active gluon to couple to the $q{\bar q}$ pair directly, the
final state gluon connected to the hard quark propagator can not be
on-shell, and we get an extra suppression of the cross
section. Such a ``symmetry breaking'' in the $q{\bar q}g$ system
does not allow us to reduce it to a symmetric gluonic $gg$ dipole
and consider its soft scattering in the same way as $q{\bar q}$
scattering. 

\begin{figure}[tb]
 \centerline{\includegraphics[width=0.35\textwidth]{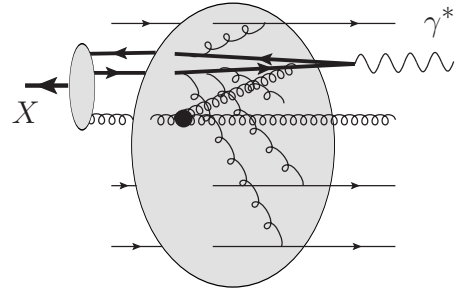}}
   \caption{\label{fig:qqgsat}
   Illustration of the $q{\bar q}g$ contribution to diffractive DIS.}
\end{figure}

The corresponding physical situation is
illustrated in Fig.~\ref{fig:qqgsat}. The hard virtual
photon first fluctuates into a virtual $q{\bar q}$ pair, and the
leading configuration is when one quark (antiquark) takes most
of the photon virtuality whereas the other one is almost on-shell. Then
the most virtual quark (antiquark) emits a (less virtual) gluon, which
interacts with a slightly virtual sea gluon from the
proton background field. This last interaction produces an 
essentially on-shell final state
gluon, which contributes to the final
$X$ system. After the first hard gluon exchange both quarks have
similar and small virtualities and scatter off the
proton background field.

In order to calculate the $q{\bar q}g$ contribution to the
diffractive structure function we include a DGLAP splitting of 
the hard gluon (with longitudinal momentum fraction $x_P$) into
two gluons --- one carries momentum fraction $z_gx_P$ and
couples to the hard part, and one is on-shell and contributes to the final
state in $\gamma^*p$ c.m.s.\ frame as shown in Fig.~\ref{fig:qqgsat}. 
The diffractive structure function corresponding to the $q{\bar q}g$
contribution can be then written as (see e.g.~\cite{Ellis:1991qj})
\begin{eqnarray} \label{FDqqg}
x_PF_{q{\bar q}g}^{D(4)}\simeq
\frac{1}{N_c^2}\int\frac{dt_g dz_g}{t_g+m^2_g}\,\hat{P}_{gg}(z_g)\frac{\alpha_s(t_g)}{2\pi}
x_PF_{q{\bar q}}^{D(4)}\;
\end{eqnarray}
where the integral is regulated in the infrared by the effective gluon mass 
$m_g\simeq\Lambda_\text{QCD}$ in the gluon propagator. The factor $1/N_c^2$ is
due to averaging over the color indices (in the large $N_c$ limit)
of the extra gluon contributing
to the color singlet $X$, and
$P_{gg}(z_g)$ is the gluon--gluon splitting function
\begin{eqnarray} \label{Pgg}
\hat{P}_{gg}(z)=C_A\left[\frac{1-z}{z}+\frac{z}{1-z}+z(1-z)\right]
\end{eqnarray}
Since the $q{\bar q}$ contribution is dominated by transverse
photon polarization, in our formulation the same is true of the $q\bar qg$
contribution.

One could also include more gluons in the final state by applying DGLAP evolution 
of the gluon density, and partially
populate the rapidity gap by extra hadronic activity from the
hadronization of gluons emitted from the hard gluon in the same
way as in Monte Carlo simulations. This would lead to a
model describing a smooth transition between diffractive and
non-diffractive final states.

Let us finally comment on another approach to resumming multi-gluon exchange, which results in similar eikonal factors $[1-\exp(\dots)]$ in the amplitudes. This approach was developed by Hautmann, Kunszt, and Soper~\cite{HKS} (HKS) and by Hautmann and Soper~\cite{HS} (HS), and is applicable in both inclusive and diffractive DIS. This approach is similar to ours, employing factorization and resummation of soft $t$-channel gluons. In diffractive DIS, the incoming partonic dipole is assumed to move closely together in the transverse plane before interacting with the color field of the proton. In the HKS/HS approach all exchanged gluons are treated on the same footing. These gluons collectively carry color singlet charge and are resummed using a Wilson line. In our approach, we use conventional $k_t$-factorization in terms of the unintegrated gluon distribution, and we additionally factorize the ``hard'' gluon, which carries most of the momentum, from the rest of the exchanged gluons, which are much softer ($x'\ll x_P$) and are resummed to all orders. The resummed gluons collectively carry color octet charge, which combined with the first gluon is required to form an overall color singlet exchange. It would be interesting to examine the connections between the two approaches further.

\section{Numerical results}\label{numresults}

\begin{figure*}[tbh]
 \centerline{\includegraphics[width=1.0\textwidth,clip=]{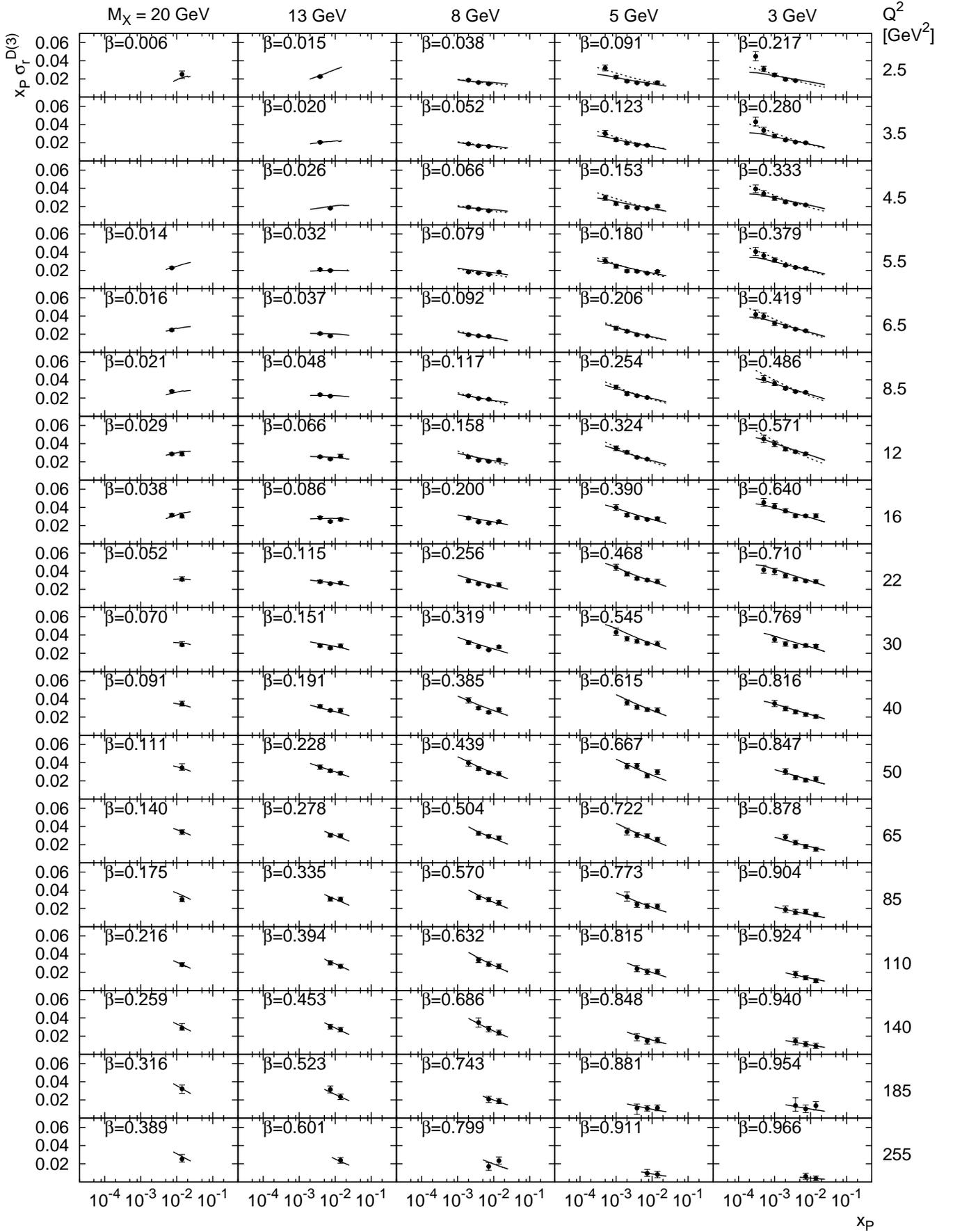}}
   \caption{\label{fig:data}
   The reduced cross section $x_P\sigma_r^{D(3)}(x_P,\beta,Q^2)$ as a function of $x_P$ for different values of $M_X$ and $Q^2$. The latest ZEUS data \cite{Chekanov:2008fh}, from diffractive deep inelastic scattering events with a large rapidity gap, compared with our model using for the gluon density in the proton the CTEQ6L1 (full line) parametrization \cite{Pumplin:2002vw} and at low $x$ and $Q^2$ also the GRV94 (dotted line) parametrization.}
\end{figure*}

The HERA data~\cite{Chekanov:2008fh,ddisexp} on DDIS are given in the form of the reduced cross
section
 \begin{equation}\label{eq-sigma}
 x_P\sigma_r^{D(3)} = x_PF_{q{\bar q},T}^{D(3)}+
 \frac{2-2y}{2-2y+y^2}\,x_PF_{q{\bar q},L}^{D(3)}+x_PF_{q{\bar q}g}^{D(3)}
 \end{equation}
expressed in terms of the diffractive structure functions $F^{D(3)}_{L,T}(x_P,Q^2,\beta)$.
The momentum transfer $t$ is integrated over since in most of the data the leading proton is not observed, and diffraction is equivalently defined through a large rapidity gap. The kinematical variable $y=Q^2/(sx_B)\leq 1$, 
where $\sqrt{s}=318$~GeV is the center-of-mass energy of $ep$-collisions in HERA. 
In Fig.~\ref{fig:data} we compare the latest ZEUS data \cite{Chekanov:2008fh}
with the numerical evaluation of our model. 
A generally very good agreement is found, but this needs to be discussed in detail in order to gain understanding of the dynamics involved. 

As discussed in Sec.~\ref{sec-GPD}, we need the generalized gluon distribution function in the proton, and use the prescription in Eq.~(\ref{ugdfsat}) for the UGDF. This reduces the problem to an input of a standard parametrization of the gluon density in the proton, \ie\ $xg(x,\mu_F^2)$. Here we mainly use the recent CTEQ6L1 parametrization \cite{Pumplin:2002vw}, which is in leading order and thereby consistent with our treatment. Below we also consider other parametrizations to illustrate the uncertainty at very small $x$ and factorization scales $\mu_F$. 
The minimum factorization scale $\mu_F$ is fixed to be $\mu^2_{F,\text{min}}=0.2\,\GeV^2$ giving rise to a minimum possible fraction of the quark longitudinal momentum $z_\text{min}$ in the phase space integral. 

The physical parameters that are fixed, are the ``soft'' coupling 
$\alpha_s^{\text{soft}}={\cal A}_1(\Lambda_\text{QCD})\simeq 0.7$, obtained from infrared-finite analytic perturbation theory (see Sec.~\ref{softpart}) and used for the coupling of the soft screening gluons, and the gluon mass $m_g\simeq \mu_{\text{soft}}\simeq \Lambda_\text{QCD}$ adopted as the infrared regulator in the gluon propagator in Eq.~(\ref{FDqqg}).

Fixing $\alpha_s^{\text{soft}}$ and $m_g$, the only free parameters in our model are $m_q^{\text{eff}}$ and ${\bar R}_g$, representing different soft effects that cannot be calculated or safely estimated. 
The constituent quark mass $m_q^{\text{eff}}$, which enters the kinematics in Sec.\ III, accounts for the soft gluon radiation from the $q{\bar q}$ dipole and corresponds to forming dressed quarks before hadronization. The soft part ${\bar R}_g$ of the off-diagonal UGDF in Eq.~(\ref{ugdfsat})
can be identified with the square root of the ``soft'' collinear PDF defined at some $x'\ll x_P$ and $\mu_{\text{soft}}$ which represents the soft scale of the color screening gluons.
The sensitivity to these parameters is discussed in the following. 

\begin{figure*}[tbh]
 \centerline{
\includegraphics[width=0.33\textwidth]{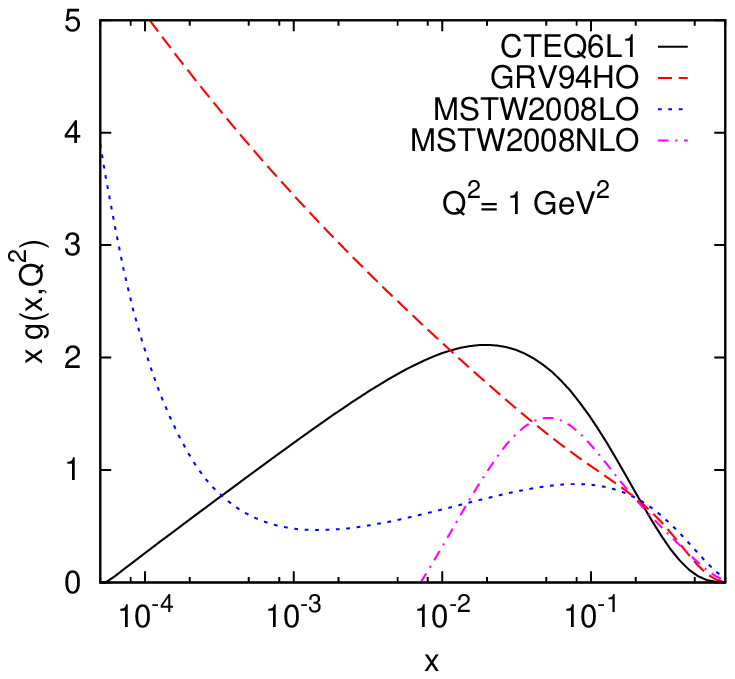}
\includegraphics[width=0.33\textwidth]{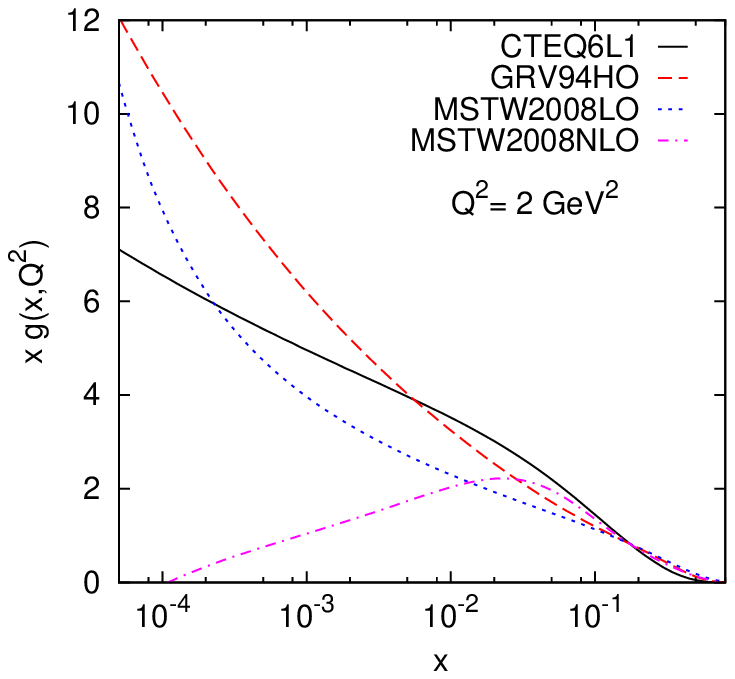}
\includegraphics[width=0.33\textwidth]{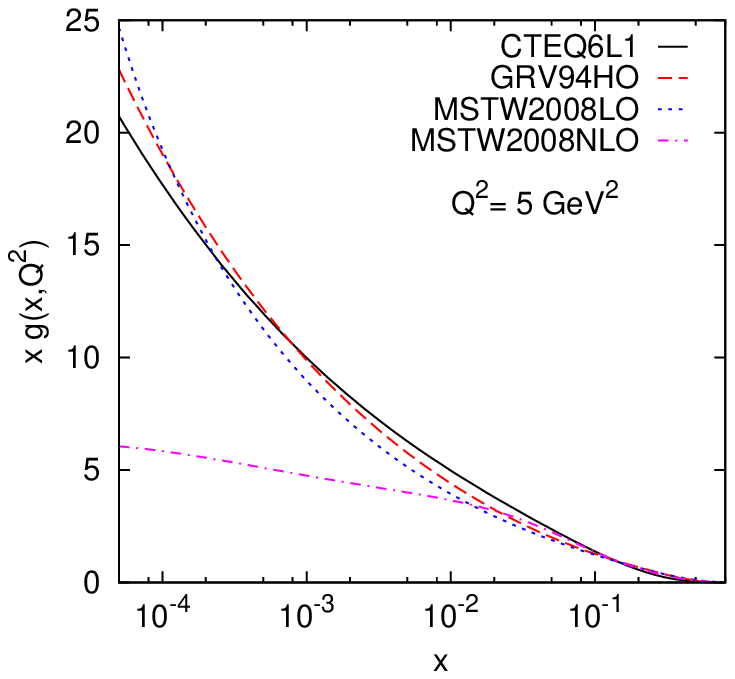}
}
   \caption{\label{fig:pdfs}
   The dependence on momentum fraction $x$ for the gluon density in the proton at the scales $Q^2=1$, 2 and 5 GeV$^2$ given by the standard parametrizations CTEQ6L1~\cite{Pumplin:2002vw}, GRV94HO~\cite{Gluck:1994uf}, MSTW2008LO, and MSTW2008NLO~\cite{MSTW}.}
\end{figure*}

The shapes of the model curves are in quite good agreement with the data,
except for a few points at extremely small $x_P\lesssim 5\times
10^{-4}$, and small scales $Q^2\lesssim 5\,\GeV^2$ and $M_X$ (see the
upper right corner of Fig.~\ref{fig:data}). 
Here, we are in the kinematical domain where the uncertainties in the parametrizations of the gluon density of the proton become extremely large, as illustrated in Fig.~\ref{fig:pdfs}. 
As can be seen, for $x\lesssim 10^{-3}$ there are substantial differences between the different gluon parametrizations and the differences become huge for $x\sim 10^{-4}$ and $Q^2\lesssim 1$ GeV$^2$. The reason is that there is no data from inclusive DIS or other processes that can measure the gluon density directly in this domain. 
This gluon PDF uncertainty strongly affects the calculated diffractive structure function at small quark fractions $z$ and/or small $Q^2$ and $M_X$, where $\mu_F$ may drop below $1$ GeV.
In this case $x'\sim x_P$, due to Eq.~(\ref{xpr}), so our basic assumptions and QCD factorization itself become less reliable.
In principle, the diffractive DIS data can be utilized for selecting the best gluon parametrization among those available in the literature or, even better, for making new gluon  parametrizations including the data on DDIS, which depend directly on the gluon density.

\begin{figure}[b]
 \centerline{\includegraphics[width=0.4\textwidth]{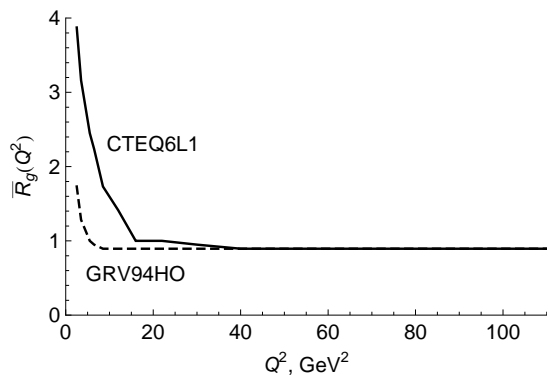}}
   \caption{\label{fig:Rg}
   $Q^2$-dependence of the normalization parameter $\bar{R}_g$, see Eq.~(\ref{ugdfsat}), 
   extracted from the HERA data, with solid curve obtained using CTEQ6L1 
   PDF~\cite{Pumplin:2002vw} and dashed curve using GRV94HO PDF~\cite{Gluck:1994uf} 
   for the gluon density in the proton.}
\end{figure}

Another signature of such uncertainties is the $Q^2$ behavior of the
soft part of the UGDF, \ie\ ${\bar R}_g(Q^2)$, which is shown in Fig.~\ref{fig:Rg} for
different PDFs. As can be seen, using either the typical leading order CTEQ6L1 PDF~\cite{Pumplin:2002vw}, which decreases at small $x_P$ and $\mu_F$, or the more regular but older GRV94HO PDF~\cite{Gluck:1994uf} (see comparison in Fig.~\ref{fig:pdfs}) to perform the fit of our model to ZEUS data results in quite different fitted ${\bar R}_g(Q^2)$. At higher scales, $Q^2\gtrsim 16\,\GeV^2$, the soft factor is quite stable at ${\bar R}_g\simeq 1$.
However, at $Q^2\lesssim 5\,\GeV^2$, the diffractive cross section calculated with CTEQ6L1 is underestimated by almost an order of magnitude, and
in order to get the correct normalization the fitted ${\bar R}_g$ value grows significantly. This is mostly because of the strong suppression in CTEQ6L1 at small $x_P\lesssim 5\times 10^{-4}$ at
scales $\mu_F^2\sim 1\,\GeV^2$. In contrast, the fit with GRV94HO,
which does not decrease at small $x_P$, leads to a more stable
behavior at low $Q^2$, such that ${\bar R}_g$ essentially becomes an overall normalization constant close to unity.

\begin{figure*}[tbh]
 \centerline{\includegraphics[width=0.9\textwidth, clip=]{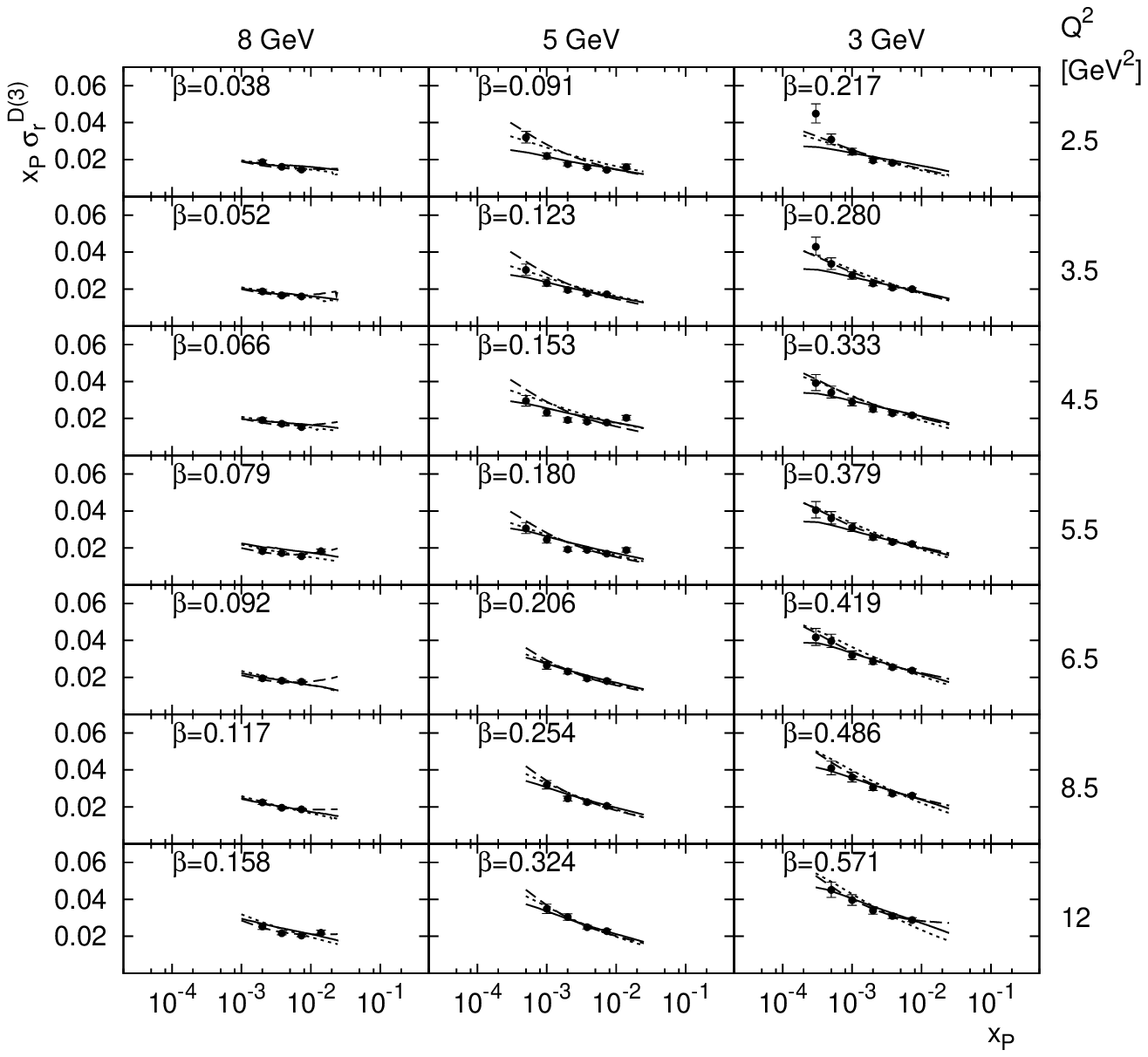}}
   \caption{\label{fig:pdfvsdata}
     The reduced cross section $x_P\sigma_r^{D(3)}(x_P,\beta,Q^2)$ as a function of $x_P$ 
     in the region of low scales $M_X\leq 3,\,5,\,8$ GeV and $Q^2<12\,\GeV^2$, 
     where the lowest $x_P$-values are reached. 
     ZEUS data \cite{Chekanov:2008fh} are compared with the results of the model 
     using CTEQ6L1 (solid line) and GRV94HO (dotted line) in the ``square root''
     UGDF prescription (\ref{ugdfsat}), 
     and with CTEQ6L1 in the ``$R_g$'' prescription (\ref{ugdfrg}) (dashed line, normalization adjusted for easy comparison with the other curves).}
\end{figure*}

In order to illustrate how uncertainties in the PDFs and in the UGDF
prescriptions affect the $x_P$-dependence in comparison with data, we compare our model 
to the data using both CTEQ6L1 and GRV94HO in the UGDF prescription of Eq.~(\ref{ugdfsat}). This is our ``normal'' prescription that gives a linear dependence of the cross section on the gluon density. For comparison we also use
CTEQ6L1 with the ``old $R_g$''-prescription defined in Eq.~(\ref{ugdfrg}), which makes the cross section depend on the square of the gluon density. 
Fig.~\ref{fig:pdfvsdata} shows the results in the bins of interest with
small scales $M_X$ and $Q^2$.
The ``old $R_g$''-prescription leads to an order of
magnitude too small diffractive structure function at all
$Q^2$, which cannot be explained by the expected normalization factor
$R_g$ of order unity in Eq.~(\ref{rg}). The corresponding curves in Fig.~\ref{fig:pdfvsdata} have therefore been normalized in order to compare the $x_P$-slopes. These slopes are in reasonable agreement with the data, but there is a tendency for a too large curvature generated by the squared gluon density, in particular at large $Q^2$.

For the curves with linear gluon density, the curves with GRV94HO lead to 
better slopes at the smallest $Q^2$ and $M_X$ than those with CTEQ6L1, but at higher scales 
they become too steep and cannot describe data. This is not surprising since the old GRV parametrization from 1994 does not take into account later data from HERA and elsewhere, but it provides an interesting alternative due to its more regular behavior at very small $x$ at low scales.   
The curves fitted with the recent CTEQ6L1 parametrization have better $x_P$-slopes at higher scales and this is therefore the main alternative in Fig.~\ref{fig:data}, in spite of its shortcoming at the very lowest $x_P$ points. 

\begin{figure*}[tbh]
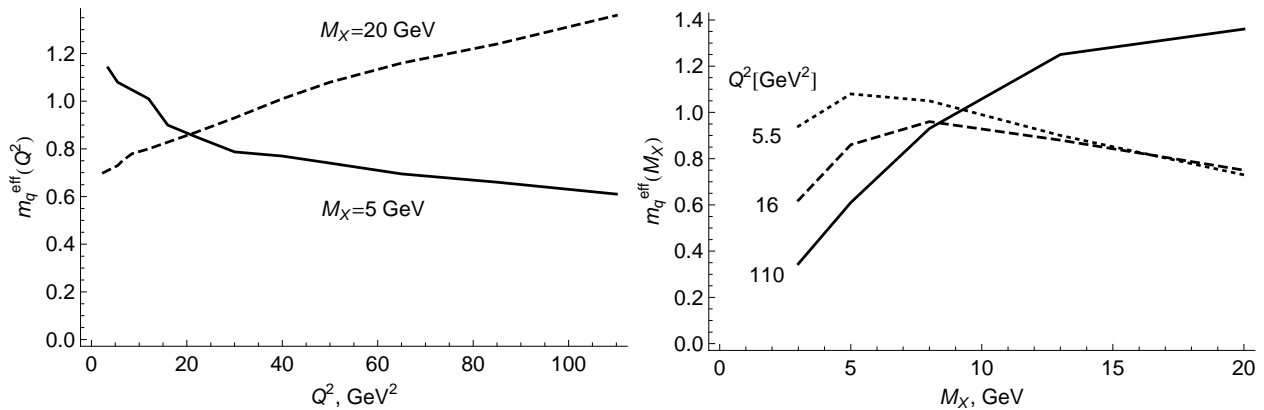

 \centerline{\includegraphics[width=0.46\textwidth]{meffQ2.eps}
\includegraphics[width=0.46\textwidth]{meffMX.eps}}
   \caption{\label{fig:meff}
   The effective quark mass parameter $m_q^{\text{eff}}$
   extracted from the fits of the HERA data shown in Fig.~\ref{fig:data}, 
   using the CTEQ6L1 gluon parametrization in the ``square root'' prescription (\ref{ugdfsat}). 
   In the left panel, as function of $Q^2$ with curves for $M_X=20$ GeV (dashed) and $M_X=5$ GeV (solid). 
   In the right panel, as function of $M_X$  with curves for $Q^2=5.5\,\GeV^2$ (dotted), $Q^2=16\,\GeV^2$ (dashed) and $Q^2=110\,\GeV^2$ (solid). }
\end{figure*}

The remaining free parameter to discuss is $m_q^{\text{eff}}$, the effective mass of the quark and antiquark in the $X$-system which is used in kinematical relations. 
In Fig.~\ref{fig:meff} we show fitted values of $m_q^{\text{eff}}$ 
at different scales $M_X$ and $Q^2$. The diffractive cross section itself is not very sensitive to $m_q^{\text{eff}}$, which therefore only varies within the physically reasonable
interval $\Lambda_\text{QCD}\lesssim m_q^{\mathrm{eff}}\lesssim1.3$ GeV. 
Thus, it is mostly of nonperturbative nature and can be interpreted as a
constituent quark mass.
Nevertheless, $m_q^{\text{eff}}$ depends on both $Q^2$ and $M_X$, indicating that both scales contribute to generating softer gluon radiation.  This dependence is, however, non-trivial. 

Indeed, a larger invariant mass $M_X$ provides a larger
phase space, which may accumulate more soft collinear gluons, leading to a
larger effective quark mass. On the other hand, a harder photon
(large $Q^2$) can probe a quark at smaller distances, so the
$Q^2$-dependence
of the quark mass obeys renormalization group evolution, i.e.\
$m_q^{\text{eff}}$ should decrease at larger $Q^2$. These two effects
are indeed observed in the description of data (see Fig.~(\ref{fig:meff})
for $M_X\lesssim 10$ GeV. At larger $M_X$ the situation changes
somewhat  due to more hard gluon radiation contributing to $M_X$
(in particular, the $q{\bar q}g$ contribution becomes important).

We now investigate the role of the different contributions to $x_P\sigma_r^{D(3)}$ in Eq.~(\ref{eq-sigma}), \ie\  $q{\bar q}$ from longitudinally and transversely polarized photons and the $q{\bar q}g$ contribution. In our results shown in Figs.~\ref{fig:data} and~\ref{fig:pdfvsdata} above, they are all included. 
We find, however, that the leading order $q{\bar q}$-dipole contribution dominates in all bins of $M_X$ and $Q^2$ and is enough to describe all data for $\beta\gtrsim 0.2$, below which the $q{\bar q}g$ contribution becomes significant and can be approximated with its leading part calculated via DGLAP splitting of the first, hard gluon. 

\begin{figure*}[tbh]
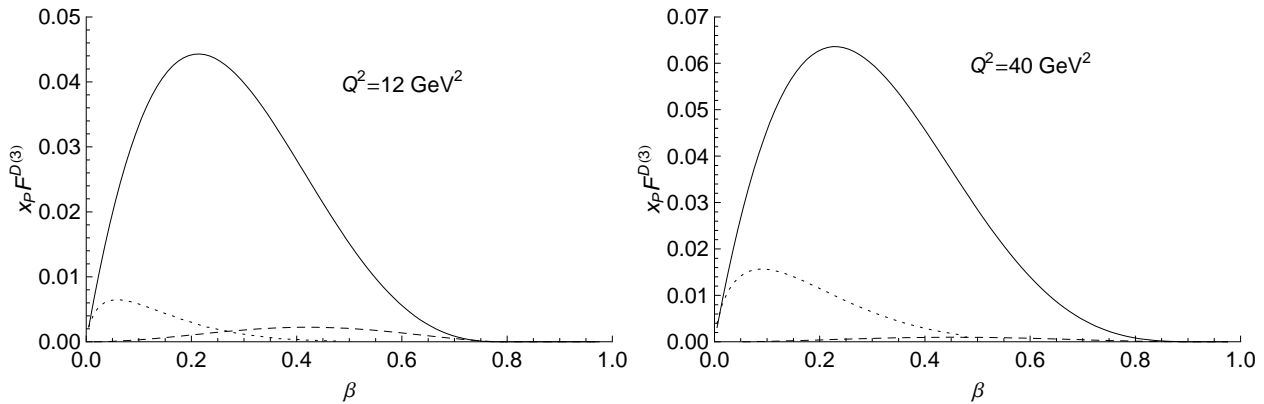

 \centerline{\includegraphics[width=0.46\textwidth]{beta12GeV.eps}
 \includegraphics[width=0.46\textwidth]{beta40GeV.eps}}
   \caption{\label{fig:bspecr}
   $\beta$-dependence of the different contributions to the diffractive structure function, 
   Eq.~(\ref{eq-sigma}), at $Q^2=12\,\GeV^2$ (left) and $Q^2=40\,\GeV^2$ (right).
   Solid lines represent the transverse part $x_PF_{q{\bar q},T}^{D(3)}$,
   dashed lines the longitudinal part $x_PF_{q{\bar q},L}^{D(3)}$
   and dotted lines the gluonic $q{\bar q}g$ contribution $x_PF_{q{\bar q}g}^{D(3)}$. 
   As usual, the prescription of Eq.~(\ref{ugdfsat}) for the unintegrated gluon density is used 
   with the  CTEQ6L1 parametrization. (The quark effective mass is here fixed at
   $m_q^{\text{eff}}=0.9$ GeV, being the mean value at these scales.)}
\end{figure*}

Fig.~\ref{fig:bspecr} shows the $\beta$-spectra for the different contributions.
They all vanish in the limits of $\beta\to 0$, corresponding to large $M_X$, and $\beta\to 1$, corresponding to small or vanishing $M_X$ with production of resonances, which is not taken into account here, or no available phase space. The transverse $q{\bar q}$ contribution dominates over the other contributions. The longitudinal $q{\bar q}$ contribution is always small, although it becomes slightly larger at smaller $Q^2$ scales. The gluonic $q{\bar q}g$ contribution becomes relatively larger both at high $Q^2$ and small $\beta$, where it gives an important contribution that must be taken into account.

\section{Conclusions and outlook}

We have in this paper developed a proper QCD framework for diffractive hard scattering, 
which contains both hard and soft dynamics. The hard part produces a well-defined state of emerging partons, and  
the soft part is the rescattering of these partons with the color field of the proton remnant. 
We have demonstrated that, by taking the Fourier transform from momentum space to impact parameter space, 
the overall amplitude can be factorized into separate amplitudes for these hard and soft parts. 
This provides a substantial simplification for the calculation and is consistent with the physical insight that soft, 
long distance processes cannot affect the hard process occurring on a short distance scale. 

The hard part is calculated using perturbative QCD, in the same way as for inclusive DIS. A perturbative hard scale is provided by the photon virtuality $Q^2$ and invariant mass $M_X$ of the diffractive system, and the process thus occurs at a space-time scale much smaller than the proton size. For small $x$, the hard subprocess 
$\gamma^* g\to q\bar{q}$ dominates. This process is mediated by a single gluon exchange 
taking most of the longitudinal momentum transfer, and leaves a proton remnant 
consisting of the three valence quarks in a color octet state. 
The proton remnant carries most of the beam momentum, and is therefore well
separated in rapidity from the $q\bar{q}$ system. 

The soft part of the amplitude accounts for the rescattering of the $q\bar{q}$ pair (in a color octet state) 
with this remnant. This rescattering is dominated by multiple exchanges of soft gluons, which have larger couplings and 
less propagator suppression. The result is a negligible change of the momenta of the emerging partons, 
but an important change of phase is picked up --- this is the essence of the eikonal approximation.
We find that summing over an arbitrary number of exchanged gluons leads to exponentiation of the soft amplitude, 
which can be written in a closed analytic form free of infrared divergences. The color exchange, treated 
in the large-$N_c$ approximation, leads to an overall color singlet exchange between 
the $q{\bar q}$ dipole and the proton. These two color singlet systems then hadronize independently separated by 
a gap in rapidity as characteristic signature of diffractive scattering. 

By invoking physical considerations based on the uncertainty principle, which limits the possible resolution 
of small momentum transfers, we obtain simplifications of the otherwise complicated angular relations in the 
impact parameter space. In essence, the orientation of the $q\bar{q}$-dipole relative to the proton color field 
is physically irrelevant and can be averaged out. 

In addition to the leading order contribution from the $q\bar{q}$-dipole, we have also included the next-to-leading 
contribution $q\bar{q}g$ with an extra gluon in the final state. Here, we find that the most important contribution 
is emission of this gluon from the exchanged hard gluon (in $\gamma^* g\to q\bar{q}$), which can be well 
approximated by leading logarithmic DGLAP emission. 

Numerical evaluation of the analytical results gives good agreement with the precise HERA data on 
the diffractive deep inelastic cross section. The $q\bar{q}$ contribution is indeed dominant, 
but at $\beta \lesssim 0.2$, the $q\bar{q}g$ contribution is important. 
At very small $x_P\lesssim 5\times 10^{-4}$ and scales $\mu_F^2\sim 1$ GeV$^2$ the gluon density in the proton, 
which is used as input in our calculation, is very poorly known and gives a complication in the comparison with
the few HERA data points in this extreme region. Standard up-to-date parametrizations have a too low gluon density
in this $x,\mu_F^2$ region, whereas, e.g., the old GRV94 gluon density does better. Since the diffractive 
cross section depends directly on the gluon density, and not only indirectly via DGLAP evolution as for inclusive DIS, 
one here obtains an interesting possibility to constrain the gluon density at very small $x$.  

Having demonstrated that our theoretical formalism for DDIS does describe HERA data, 
one may then extract the part describing the multigluon exchange process and apply it to other 
hard scattering processes. This soft rescattering description ought to be universal, due to 
the factorization of the hard and soft amplitudes. Thus, one may apply it together with hard processes 
in $p\bar{p}$ collisions at the Tevatron to describe the different hard diffractive processed observed there, and then 
go to the higher energies at the LHC. One may also apply it to more detailed observables in diffractive DIS, such as diffractive dijets or diffractive vector meson production.

However, not only diffractive processes are of interest. The Soft Color Interaction (SCI) model discussed above has previously been successfully applied to both charmonium production and $B$-meson decays~\cite{charmsci}, and one may expect the model presented in this paper to have interesting applications in such processes too. 
Moreover, the multi-gluon exchange mechanism will also affect the underlying event, since it effectuates 
color exchanges that modify the color-string topology and thereby the hadronic final state after hadronization.
The underlying event is important in its own right to understand non-perturbative QCD dynamics, and also for  
understanding of inclusive events when subtracting the Standard Model background in searches for new phenomena at LHC.   

Finally we note that in deriving the theoretical formalism presented here, we have {\em not} used any assumptions 
or results from the previous Soft Color Interaction (SCI) model. Our new formalism stands on its own, based on QCD theory 
and basic physical arguments. The formalism can, however, explain why the simple SCI
model has been so successful in describing data on diffractive hard scattering and other phenomena. 
The assumptions of the SCI model 
as well as its major features are essentially what comes out as results of the present paper. 
Of course, our new formalism has a richer dynamical structure and we will therefore attempt to improve the 
Monte Carlo implementation of the SCI model by replacing its fixed probability for soft gluon exchanges with a 
mechanism based on the above amplitude for the multiple soft gluon exchanges. This will introduce a non-trivial 
dependence on the kinematical variables, giving a new level of event-to-event variations. As usual with full event 
simulation using Monte Carlo, this will give access to more detailed studies of both the employed theoretical 
model and its comparison to data in terms of the indicated more elaborate observables.


\centerline{\textbf{Acknowledgments}}

This work was supported by the Swedish Research Council and the Carl
Trygger Foundation. We are grateful to Igor Anikin for valuable
discussions.

\end{document}